\documentclass[final,times,twocolumn,nopreprint]{elsarticle}

\usepackage{amssymb}
\usepackage{latexsym}

\usepackage{amsthm}
\usepackage{float}
\usepackage{algorithm}
\usepackage{algpseudocode}
\usepackage{multirow}
\usepackage{graphicx}
\usepackage{booktabs}
\usepackage[]{graphbox}
\usepackage{array}
\usepackage{url}
\usepackage{subfigure}
\usepackage{tikz}
\usetikzlibrary{shadows,arrows,shapes,patterns,fit,positioning}
\pgfdeclarelayer{background}
\pgfdeclarelayer{foreground}
\pgfsetlayers{background,main,foreground}
\pgfdeclarepatternformonly{south west lines}{\pgfqpoint{-0pt}{-0pt}}{\pgfqpoint{3pt}{3pt}}{\pgfqpoint{3pt}{3pt}}{
        \pgfsetlinewidth{0.4pt}
        \pgfpathmoveto{\pgfqpoint{0pt}{0pt}}
        \pgfpathlineto{\pgfqpoint{3pt}{3pt}}
        \pgfpathmoveto{\pgfqpoint{2.8pt}{-.2pt}}
        \pgfpathlineto{\pgfqpoint{3.2pt}{.2pt}}
        \pgfpathmoveto{\pgfqpoint{-.2pt}{2.8pt}}
        \pgfpathlineto{\pgfqpoint{.2pt}{3.2pt}}
        \pgfusepath{stroke}}

\usepackage{url}
\usepackage{xcolor}

\usepackage{hyperref}

\definecolor{newcolor}{rgb}{.8,.349,.1}

\journal{Medical Image Analysis}

\begin{document}


\begin{frontmatter}

\title{Low-field magnetic resonance image enhancement via stochastic image quality transfer}

\author[1,2,3]{Hongxiang Lin\corref{cor1}}
\cortext[cor1]{Corresponding author.}
\ead{hxlin@zhejianglab.edu.cn, harry.lin@ucl.ac.uk}
\author[2,3]{Matteo Figini}
\author[4]{Felice D'Arco}
\author[5]{Godwin Ogbole}
\author[6]{Ryutaro Tanno}
\author[2,3,7]{Stefano B. Blumberg}
\author[2,3]{Lisa Ronan}
\author[8]{Biobele J. Brown}
\author[9,10]{David W. Carmichael}
\author[8]{Ikeoluwa Lagunju}
\author[10]{Judith Helen Cross}
\author[3,8]{Delmiro Fernandez-Reyes}
\author[2,3]{Daniel C. Alexander}

\address[1]{Research Center for Healthcare Data Science, Zhejiang Lab, Hangzhou 311121, Zhejiang, China}
\address[2]{Centre for Medical Image Computing, University College London, London WC1E 6BT, United Kingdom}
\address[3]{Department of Computer Science, University College London, London WC1E 6BT, United Kingdom}
\address[4]{Department of Radiology, Great Ormond Street Hospital for Children, London WC1N 3JH, United Kingdom}
\address[5]{Department of Radiology, College of Medicine, University of Ibadan, Ibadan 200284, Nigeria}
\address[6]{Google DeepMind, London N1C 4AG, United Kingdom}
\address[7]{Centre for Artificial Intelligence, University College London, London WC1E 6BT, United Kingdom}
\address[8]{Department of Paediatrics, College of Medicine, University of Ibadan, Ibadan 200284, Nigeria}
\address[9]{School of Biomedical Engineering \& Imaging Sciences, King's College London, London NW3 3ES, United Kingdom}
\address[10]{UCL Great Ormond Street Institute of Child Health, London WC1N 3JH, United Kingdom}


\begin{abstract}
Low-field ($<1$T) magnetic resonance imaging (MRI) scanners remain in widespread use in low- and middle-income countries (LMICs) and are commonly used for some applications in higher income countries e.g. for small child patients with obesity, claustrophobia, implants, or tattoos. However, low-field MR images commonly have lower resolution and poorer contrast than images from high field (1.5T, 3T, and above). Here, we present Image Quality Transfer (IQT) to enhance low-field structural MRI by estimating from a low-field image the image we would have obtained from the same subject at high field. Our approach uses (i) a stochastic low-field image simulator as the forward model to capture uncertainty and variation in the contrast of low-field images corresponding to a particular high-field image, and (ii) an anisotropic U-Net variant specifically designed for the IQT inverse problem. We evaluate the proposed algorithm both in simulation and using multi-contrast (T1-weighted, T2-weighted, and fluid attenuated inversion recovery (FLAIR)) clinical low-field MRI data from an LMIC hospital.  We show the efficacy of IQT in improving contrast and resolution of low-field MR images. We demonstrate that IQT-enhanced images have potential for enhancing visualisation of anatomical structures and pathological lesions of clinical relevance from the perspective of radiologists. IQT is proved to have capability of boosting the diagnostic value of low-field MRI, especially in low-resource settings. 
\end{abstract}


\end{frontmatter}


\section{Introduction}
\label{sec:introduction}
Magnetic Resonance Imaging (MRI) is  ubiquitous in neurology and many other areas of medicine. While high-field scanners, typically 1.5T and 3T, are the current clinical standard in high income countries (HICs), low-field scanners, less than 1T, remain widespread in many low and middle income countries (LMICs), due to ease of installation, affordability, and robustness to power outages. However, low-field images lack the diagnostic information content of high field, because they have lower signal-to-noise ratio (SNR) and may have lower contrast-to-noise ratio (CNR) at equivalent spatial resolution. For example, grey matter (GM) / white matter (WM) contrast on T1-weighted images is usually lower than at high field even at equivalent SNR and spatial resolution, because the longitudinal relaxation time (T1 coefficient) of the two tissues are more similar below 1T~\citep{Marques2019} than above~\citep{Jones2018}.  


Image Quality Transfer (IQT)~\citep{Alexander2014,Alexander2017,Blumberg2018,lin2019deep,Tanno2021,Lin2021b} is a machine learning framework used to enhance low-quality clinical data, e.g. from a rapid acquisition protocol and/or standard hospital scanner, with more abundant information in high-quality images, e.g. from rich acquisition protocols too lengthy to run on every patient and/or bespoke experimental scanners available only in specialist centres. Early work~\citep{Alexander2017,Blumberg2018,Tanno2021} focused on diffusion MRI and showed compelling ability to enhance both contrast and spatial resolution, for example enabling tractography to recover small pathways impossible to reconstruct at the acquired resolution. The IQT technology is distinct from super-resolution in computer vision~\citep{Yang2019,Wang2021pami}, as it aims not only to increase resolution but also to enhance other features, such as image contrast. It also differs from modality transfer, see e.g.~\citet{Burgos2014,Jog2015,Wolterink2017,Cohen2018,Iglesias2021}, as it focuses on enhancing the quality (resolution and contrast) rather than content. Other researchers aim to construct ``7T-like'' images from 3T addressing a similar problem to ours. For example, \citet{Bahrami2016b,Bahrami2016a} employ first canonical correlation analysis and subsequent deep convolutional neural networks (CNNs) via a process similar to IQT. \citet{Xiang2018} extend the process with Cycle-Consistent Adversarial Networks on unpaired 3T and 7T data. \citet{Zhang2018} investigate dictionary-based sparse coding regression for synthesising 7T from 3T MRI. \citet{Kaur2019} learn intensity transformations from a single 3T image to a 7T image. \citet{qu2020} learn a deep learning model in a wavelet domain to map 3T MRI to synthesised 7T MRI. However, contrast relationships between 1.5T or 3T and low field are more complex and non-linear than they are between say 3T and 7T as both biophysical models~\citep{Brown2014,Marques2019}, empirical studies~\citep{Rooney2007,Wu2016} and the recent experiment~\citep{Liu2021} demonstrate.

Two features of IQT naturally support an application of low-field MR image enhancement. First, rather than directly using pairs of images acquired on different platforms, IQT often uses simulated low-quality data from high quality to obtain matched pairs. Indeed, it is challenging to get a large number of paired data from high and low-field scanners, but sufficient low-field examples are available to construct a forward model that approximates low-field images from high-field examples. Using simulated low-quality data has the additional advantage of avoiding confounds of misalignment, since even small spatial shifts between image pairs can strongly disrupt the learned mappings that blur output images. Although this reduces realism of the low-field training data, it generally proves sufficient to enhance real images, accepting that perfect estimation of the high-field equivalent is not practically feasible. Nevertheless, the ultimate aim in practice is to enhance the image quality rather than emulating high-field images. Second, IQT implementations generally use patch regression and learn mappings between image patches rather than whole images and piece together estimates of high-quality images by enhancing low-quality images patch by patch. This further reduces the amount of data required for training. Typically, IQT uses input patches with relatively larger field of view compared to output patches, which reduces deterioration~\citep{Innamorati2020} in the mapping by enabling it to exploit local image structure. However, this creates a trade off between data/memory requirements, which increase with patch-size, and deterioration, which decreases with patch size.  


In this paper, we aim to enhance clinical low-field MR images from LMIC healthcare settings. In contrast to standard hospital scanners in HICs, special challenges when using low-field scanners in LMIC clinics drive the choice of approach and implementation. First, to counteract the low SNR at low-field, clinical practitioners commonly adjust acquisition parameters for each scan to optimise contrast for individual patients leading to data sets with inconsistent image contrasts. Next, acquisition protocols routinely acquire only non-adjacent thick slices to reduce the acquisition time and cross-talk artifacts. Furthermore, the largely anisotropic voxels and the missing information from slice gaps disrupt standard super-resolution and contrast transfer approaches. Lastly, the practitioners may manually adjust the acquisition geometry according to position and direction of scanning head, resulting in data sets with diverse voxel spacings. 

To address these challenges, we build on the IQT framework to construct a mapping that estimates high-field images, approximating images acquired in e.g. a HIC imaging centre, from clinical low-field images, such as those acquired in LMIC hospitals. To achieve this we introduce a stochastic decimation model that produces synthetic low-field images from real high-field images and thus provides matched image-pairs for training. The model captures the variation in low-field image contrast that we observe in clinical data sets arising from variation in both patient anatomy and adjustments in imaging protocols. We further introduce a variant architecture of the super-resolution U-Net designed to enable gradual upscaling of image features and demonstrate advantages of this architecture for low-field MRI IQT. The processing pipeline further involves a histogram alignment step to harmonise training and test data sets, which proves necessary to accommodate the variation in contrast among both real and synthetic low-field images. The full pipeline refines and builds on preliminary methods reported in~\citet{lin2019deep} and~\citet{Figini2020} and we substantially extend the earlier experimental work. In particular, we train IQT mappings for multiple structural contrasts: T1-weighted (T1w), T2-weighted (T2w) and fluid-attenuated inversion recovery (FLAIR) images. We use simulations to tune parameters of the system and demonstrate the benefits of the resulting IQT system on clinical data from the University College Hospital (UCH) Ibadan, Nigeria by evaluating the impact of the IQT enhancement on visualisation of anatomic and pathological structures. The source code is available on Github: \url{https://github.com/hongxiangharry/Stochastic-IQT}.

\section{Methods}

\begin{figure*}[!t]
    \centering
    \includegraphics[width=1\textwidth]{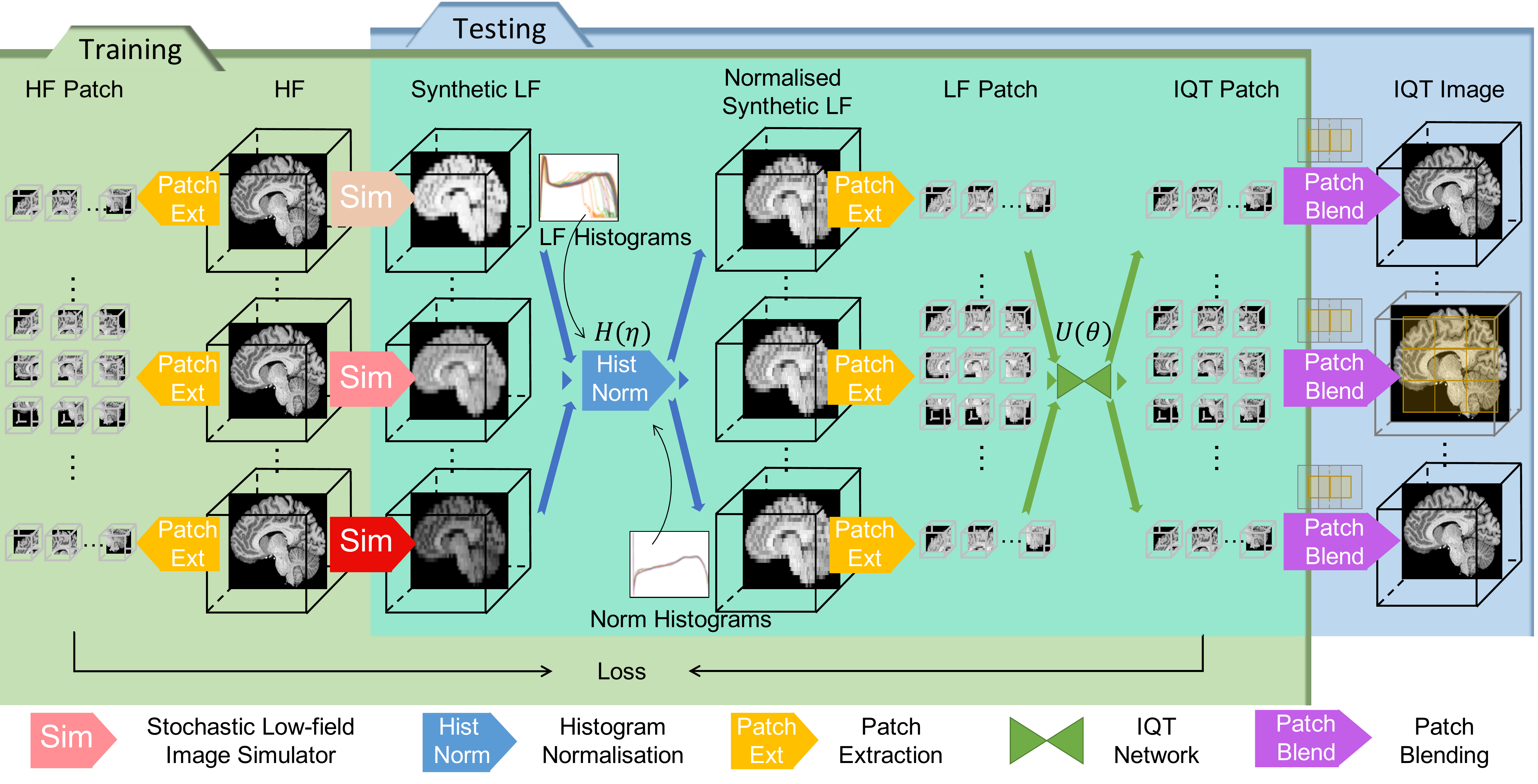}
    \caption{Schematic diagram of the Stochastic Image Quality Transfer framework. Training uses a stochastic low-field image simulator to approximate low-field (LF) from high-field (HF) images. Histogram normalisation removes gross intensity misalignments. We then extract a set of matched low- and high-field patch pairs and train a network to estimate high-field patches from low-field patches.  At test time, we input a low-field image patch by patch to estimate high-field patches, which we piece together via a patch-blending process to output the estimated high-field image.}
    \label{fig:iqt-schematic-diagram}
\end{figure*}

This section specifies the components of our implementation for low-field MRI stochastic IQT. Figure~\ref{fig:iqt-schematic-diagram} gives an overview of the full system. The key components are a stochastic low-field image simulator, a histogram normalisation module, and a patch-based deep neural network. At training time, the system first constructs a training set of matched pairs of image patches. Specifically, the stochastic low-field image simulator inputs a high-field image and outputs the corresponding synthetic low-field image each with randomly generated contrast sampled from a pre-specified \textit{a priori} distribution. This distribution reflects the tissue-specific contrast-range observed in real-world low-field images. The full set of low-field images is then used to train a histogram normaliser to remove intensity scale differences.  A sliding-window patch-extraction process then crops a set of corresponding patch-pairs from the set of paired images to obtain the final training set. Our novel anistropic U-Net uses that training set to learn a mapping from low to high-field patches by minimising patch-wise mean squared error. At test time, a new unseen low-field image first goes through the trained normaliser and then patch-by-patch through the trained network to recover an initial estimated high-field image, which undergoes a patch-blending process to produce the final output. The following subsections detail each component.

\subsection{Stochastic low-field image simulator}\label{sec:method_simulator}
We propose a stochastic low-field image simulator, used as our IQT forward model, to generate synthetic low-field images from the corresponding high-field images; see the first procedure in Figure~\ref{fig:iqt-schematic-diagram}. Algorithm~\ref{alg:1} details the procedure that simulates a synthetic low-field image $\hat{X}$ from a high-field image $Y$. In summary, we downsample the image resolution to match low-field acquisitions, allocate stochastically an SNR reduction for each tissue type, adjust the signal in each voxel according to the SNR reduction, and add noise.


We use Statistical Parametric Mapping (SPM)~\citep{Ashburner2005} to first skull strip the high-field image, and then determine probabilistic voxel masks for key tissue types (WM and GM) using the unified segmentation algorithm. The segmentation of a high-field image $Y$ obtains three tissue-category masks $M^j$, $j \in J=\{WM, GM,oth\}$, respectively for WM, GM, and all \textit{oth}er non-GM/WM tissue classes provided by the SPM's algorithm. Each mask is a probability tensor whose element $M^j(\mathbf{v}) \in [0, 1]$ represents how likely it is that the voxel coordinate denoted by $\mathbf{v}$ belongs to tissue type $j$. All the masks are used as the inputs for Algorithm~\ref{alg:1}.

The procedure downsamples $Y$ and $M^j$ ($j \in J=\{WM, GM,oth\}$) only along the slice direction (superior to inferior as $z$-direction). We convolve with a one-dimensional Gaussian filter $h_\varphi(z)= (2\varphi^2\pi)^{-1/2}\exp(-z^2/(2\varphi^2))$ in the $z$-direction with $\varphi$ chosen so that the full-width at half maximum (FWHM) equals the slice thickness in the low-field images, i.e. $\varphi=r \phi_z/\sqrt{8\ln 2}$ where $\phi_z$ is the slice thickness in the $z$-direction of the high-field images, $r \phi_z$ is that of the low-field images, and $r$ is the downsampling factor. For resampling, the distance $D_r$ between slices for the low-field image is set to be $D_r = r (\phi_z+g_z)$ where $g_z$ is the gap in the $z$-direction of the high-field image. Tissue falling in the gaps has negligible contribution to the signal in the simulated image, as in real acquisitions. Thus, the downsampled high-field image $Y_{\downarrow r}$ and the downsampled masks $M_{\downarrow r}^j$ are 
\begin{eqnarray}
   Y_{\downarrow r}(\cdot, z)= \sum_{k\in\mathbb{Z}}\delta(z-kD_r)(Y(\cdot, z) \star_z h_{\varphi}(z)), \\
   M_{\downarrow r}^j(\cdot, z)=\sum_{k\in\mathbb{Z}}\delta(z-kD_r)(M^j(\cdot, z) \star_z h_{\varphi}(z)),
\end{eqnarray}
where $\star_z$ denotes a one-dimensional convolution operation on $z$-direction and $\delta(z)$ is the Dirac delta function. 

The procedure calculates the mean high-field signal $\mu_{Y}^j$ for each tissue type $j$ in downsampled images as follows:
\begin{equation}
   \mu_{Y}^j = \frac{\sum_{\mathbf{v}}M_{\downarrow r}^j(\mathbf{v})Y_{\downarrow r}(\mathbf{v})}{\sum_{\mathbf{v}}M_{\downarrow r}^j(\mathbf{v})},
\end{equation}
and the corresponding SNR as
\begin{equation}
   \mathrm{SNR}_{Y}^j = \frac{\mu_{Y}^j}{\sigma_{Y}},
\end{equation}
where $\sigma_{Y}$ is the standard deviation of the locally sampled noisy background voxels in the high-field image $Y$. 
We sample SNRs of WM and GM in the low-field image, denoted by $\vec{a} = (\mathrm{SNR}_{X}^{WM}, \mathrm{SNR}_{X}^{GM})$, from a default bivariate probability distribution $P$, which takes various forms in Section~\ref{sec:results_training_contrast_norm} and expresses the uncertainty in the low-field image contrast. With values for each SNR or $\vec{a}$, we can evaluate ratios of low-field-to-high-field image intensity, $l^{WM}$ and $l^{GM}$, for both WM and GM respectively. We then re-scale the high-field images with the ratios of image intensity of each tissue category, recombine the tissue-specific maps, and add Gaussian random noise with variance $\sigma_{X}^2-\sigma_Y^2$ to obtain the synthetic low-field image $\hat{X}$. To summarise, the low-field image simulator generates:
\begin{equation}
    \hat{X}(\mathbf{v}) = Y_{\downarrow r}(\mathbf{v})\sum_{j\in J}l^j(\vec{a}) M_{\downarrow r}^j(\mathbf{v})  +\mathcal{N}(0, \sigma_{X}^2-\sigma_{Y}^2).
\end{equation}

\begin{algorithm}[!t]
\caption{Stochastic Low-Field Image Simulator}\label{alg:1}
\textbf{Input:} High-field (HF) image $Y$, masks $M^j$ for $j\in J=\{WM, GM, oth\}$, downsampling factor $r\in\mathbb{N}$, low-field (LF) tissue mean SNR distribution $P$, and LF and HF noise levels $\sigma_X$ and $\sigma_Y$\footnotemark.
\begin{algorithmic}[1]
\State $Y_{\downarrow r}=\sum_{k\in\mathbb{Z}}\delta(z-kD_r)(Y \star_z h_{\varphi})$; \Comment{Blur/resample along $z$.}
\State $M_{\downarrow r}^j=\sum_{k\in\mathbb{Z}}\delta(z-kD_r)(M^j\star_z h_{\varphi})$; \Comment{As Step 1 for masks.}
\State $\mu_Y^j=\frac{\sum_{\mathbf{v}}Y_{\downarrow r}^j(\mathbf{v})M_{\downarrow r}^j(\mathbf{v})}{\sum_{\mathbf{v}}M_{\downarrow r}^j(\mathbf{v})}, j=WM,GM$; \Comment{Mean signals for HF.}
\State $\vec{a}=(\mathrm{SNR}_X^{WM}, \mathrm{SNR}_{X}^{GM})\sim P$; \Comment{Sample SNRs for LF.}
\State $\mu_X^j = \mathrm{SNR}_X^j \sigma_X, j=WM,GM$; \Comment{Mean signals for LF.}
\State $l^j=\left\{\begin{array}{ll}
\mu_X^j/\mu_Y^j, & j=WM,GM, \\
    1, & oth; \\
\end{array}\right.$\Comment{Evaluate multipliers.}
\State $\hat{X}=\sum_{j\in J}l^j M^j_{\downarrow r} Y_{\downarrow r}$; \Comment{Transfer contrast.}
\State $\hat{X}(\mathbf{v})\rightarrow\hat{X}(\mathbf{v})+\mathcal{N}(0, \sigma_{X}^2-\sigma_{Y}^2)$  $\forall \mathbf{v}$. \Comment{Add noise.}
\end{algorithmic}
\textbf{Output:} Noisy synthetic LF image $\hat{X}$.
\end{algorithm}

\footnotetext{By default, one can estimate the input noise levels by $\sigma_X = \mu_Y^{WM}/\mathrm{SNR}_X^{WM}$ fixing the mean WM intensity at low field to that at high field, i.e. $\mu_X^{WM}=\mu_Y^{WM}$, and $\sigma_Y\simeq 0$ as it is negligible compared to $\sigma_X$.}

\subsection{Histogram normalisation}
The procedure in Algorithm~\ref{alg:1} provides us with a set of the $N$ matched image-pairs with $N$ different contrasts for training.  However, the range of intensity scales can vary among synthetic and real low-field images, which disrupts both training and testing application. Therefore, prior to training, we use histogram normalisation to align image intensity ranges; the same alignment adjusts intensity ranges of test images to the training range prior to application of trained models. We use Nyul's algorithm~\citep{Nyul2000} for this histogram normalisation, by first computing average histogram percentile-landmarks, denoted by $\eta$, over the set of synthetic low-field images, and then mapping all synthetic and real image histograms to have corresponding landmarks via a piecewise linear intensity transformation denoted by $H(\eta)$ for Figure~\ref{fig:iqt-schematic-diagram}.

\subsection{Training data set and patch extraction}\label{sec:method-train-data-patch-ext}
Here, we use a patch-based approach to the IQT mapping. For training this requires a set of matched high-field and low-field image patches, which we obtain by cropping corresponding high-field and normalised synthetic low-field images at corresponding regularly spaced locations. We use the sliding window technique\footnote{~Sliding window technique can be called through the function \texttt{extract\_patches} in the Python package \textit{scikit-learn 0.22}.} to extract overlapping patches. This provides training patch pairs in $\tau_M=\{(\mathcal{X}_i, \mathcal{Y}_i)|i=1,\cdots, M\}$, which contains $M$ paired patches. Each 3D low-field input patch $\mathcal{X}_i$ has size $l\times h\times d$ voxels and the corresponding high-field output patch $\mathcal{Y}_i$ has size $l\times h\times rd$ voxels. Our default $\tau_M$ uses a high-field patch size of $32\times32\times32$, and extraction step size of $16$, $16$, and $16/r$ along $x$-, $y$-, and $z$-directions, respectively, of each low-field training image paired with corresponding patch-positions in the high-field images. This provides low-field and high-field patch sizes of $32\times 32\times (32/r)$ and $32\times 32\times 32$, respectively. Patches containing $>80\%$ background voxels are excluded from the patch library. 

\subsection{Deep learning framework}\label{sec2.1}
The classical 3D isotropic U-Net~\citep{Cicek2016} maps between two image arrays (input and output of the encoder-decoder) assuming isotropic, i.e. perfectly cubic, voxels. Each level for a typical U-Net consists of several convolutional layers together with a pooling layer. The activation from each level in the encoder is concatenated to the input features to the same level in the decoder, enabling the network to integrate both local and global image features. The U-Net uses zero-padding for convolution operations to make the input feature dimensions align with the output feature dimensions after convolution.

\begin{figure*}[!t]
    \centering
    \includegraphics[width=0.95\textwidth]{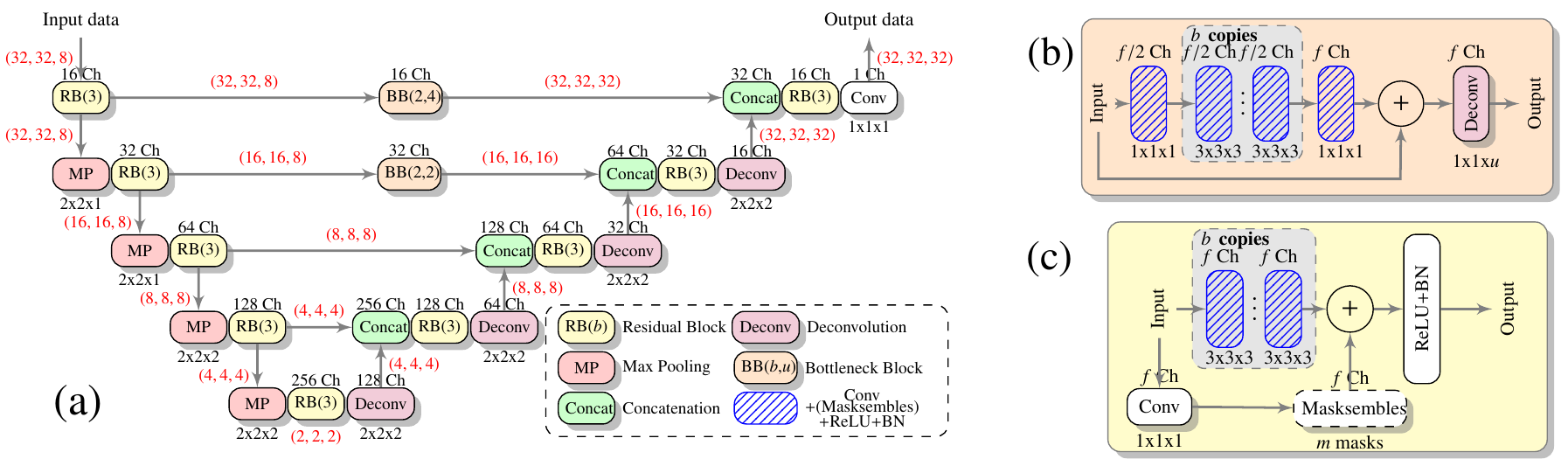}
\caption{(a) Anisotropic (ANISO) U-Net architecture (example for the super-resolution factor of $r=4$). The operations, (b) Bottleneck Block $BB(b,u)$ with $f$ filters and (c) Residual Block $RB(b)$ with $f$ filters, are specified. The round boxes correspond to the different operations illustrated in the bottom right of (a). Masksembles Layer is an optional module marked by a dashed boundary. The number of output channels, abbreviated as “Ch”, and the kernel size are denoted on top and bottom of the boxes. The arrows represent transfer of data with its corresponding shape highlighted.}
\label{fig:unet}
\end{figure*}

Here we adapt the U-Net architecture to map input and output patches that differ in voxel dimension by $r$, now denoting the upsampling factor, in the slice direction. 
Figure~\ref{fig:unet} illustrates for the case of $r=4$ where the ANISO U-Net first partially downsamples the first two dimensions until the voxels become isotropic and thereafter conducts isotropic down- and up-sampling. To achieve this, we define the following two operations:

\textbf{Bottleneck Block.} To incorporate a super-resolution transformation into the U-Net, we propose a bottleneck block  to connect corresponding levels of the contracting and expanding paths, as shown in Figure~\ref{fig:unet}(b). The design is inspired by the bottleneck block in ResNet~\citep{He2016} and Fast Super-Resolution CNN (FSRCNN)~\citep{Dong2016}. The bottleneck block has three hyperparameters: the number of input filters $f$, the number of shrinking layers $b$ and the local up-sampling scaling factor $u$. It shrinks half of the filters on consecutive $3\times 3\times 3$ convolutional layers between two endpoint convolutions with a kernel size of $1\times 1\times 1$. All convolution layers are activated by Rectified Linear Unit (ReLU) with Batch Normalization (BN). The skip connection enables the training of deeper networks~\citep{He2016}. Resolution change is efficiently carried out by a transpose convolution, or deconvolution, with the same kernel and stride of $(1,1,u)$.

\textbf{Residual Block.} To have more convolutional layers on each level, Figure~\ref{fig:unet}(c) defines the residual core, which is a revision of the residual element in~\citet{Guerrero2018}. This is a combination of several sequential $3\times 3\times 3$ convolutional layers, followed by ReLU and BN layers, skip connected with a $1\times 1\times 1$ fully convolutional layer. Then the output is attained before ReLU and BN again. Utilizing the consecutive convolutional layers enlarges each receptive field on each level.
Moreover, the appended skip connection avoids the vanishing gradient problem in neural networks with gradient-based learning methods.

\textbf{Uncertainty Quantification (UQ).} Figures~\ref{fig:unet} (a) and (c) show that we also insert a 3D variant of the Masksembles Layer~\citep{Durasov2021} adapted from its original implementation~\footnote{\url{https://github.com/nikitadurasov/masksembles}} immediately after each 3D convolutional layer to ANISO U-Net; see~\ref{appendix:uq} for details. Masksembles enables rapid UQ process by simultaneously training and testing multiple independent networks through random binary masks. Each mask is independently and randomly generated to drop out network parameters and is fixed during training and testing.

With this architecture in place, we construct the inverse model by training the convolutional deep neural network $U$. We optimise the network parameters $\theta$ by minimising the average of the pixel-wise mean squared error (MSE) over all training patch pairs $(\mathcal{X}_i,\mathcal{Y}_i), i=1,\cdots,M$ to obtain
\begin{equation}\label{loss_volume}
    \theta^* = \arg\min_{\theta}\frac{1}{M}\sum_{i=1}^M\|U(\theta; \mathcal{X}_i)-\mathcal{Y}_i\|_2^2.
\end{equation}

\subsection{Patch blending}

At test time, we reconstruct the enhanced image patch-by-patch using the sliding window technique, as described in Section~\ref{sec:method-train-data-patch-ext}. We use the simple patch-blending approach called Clipping Overlapping Patches by~\citet{Huang2018}. The window step size is selected as same as for generating the training data. Adjacent patches thus overlap and we trim each through the centre line of overlapping area and concatenate the remaining patches to produce the final output. Alternatives include Averaging Overlapping Patches and a 3D variant of Image Quilting Method~\citep{efros2001}, but empirically we find that  Clipping Overlapping Patches achieves stable and reasonable performance given sufficient training data.

\section{Experiments and results}\label{sec:experiments}

To tune and benchmark the proposed algorithm, we first run a series of experiments to evaluate the choice of network architecture against baselines and the influence of various components and parameter choices in the system in an idealised scenario using simulated data with deterministic high-field to low-field mapping, i.e. $P$ is a delta distribution.  We then consider the stochastic algorithm and, again using simulations, quantify the trade off between number of subjects and random contrasts per subject in the training data.  We also explore the effect of mismatch between the choice of $P$ to generate the training data and the $P$ that generates the test data to give insight on generalisability.  Finally, we demonstrate the algorithm on patient data acquired from UCH Ibadan and quantify radiologists' qualitative evaluation of the enhanced data in comparison to the original low-field images and corresponding high-field images.

\subsection{Data sets} 
For training, three-dimensional high-resolution T1w and T2w images were obtained from the publicly available WU-Minn Human Connectome Project (HCP) data set~\citep{Sotiropoulos2013}, acquired on a 3 Tesla Siemens Connectome Skyra scanner with a $0.7$-mm isotropic voxel. Its Repetition Time(TR)/Echo Time (TE)/Inversion Time (TI) for T1w are $2400/2.14/1000$ ms and TR/TE for T2w are $3200/565$ ms. Three-dimensional high-resolution FLAIR images were given by the “Leipzig Study for Mind-Body-Emotion Interactions” (LEMON) data set~\citep{babayan2019mind}, acquired on a 3 Tesla Siemens MAGNETOM Verio scanner with an $1.0$-mm isotropic voxel. Its TR/TE/TI are $5000/395/1800$ ms. 

We use two low-field data sets from Nigeria.  First, data set \textit{LF17} includes $39$ low-field MRI scans from neurological patients at University College Hospital (UCH) Ibadan in 2017, which we used to inform the low-field MRI simulations and to conduct qualitative analysis of our IQT algorithm. Second, \textit{LF19} consists of similar clinical low-field MRI scans from $12$ neurological patients for epilepsy surgery in 2019, which we used for both image quality analysis and radiological evaluation. Both data sets include 2D axial T1w (typically TE/TR = $15/750$ ms), T2w (typically TE/TR = $128/7500$ ms) and FLAIR (typically TE/TR/TI = $108/7500/1700$ ms) images acquired on a 0.36T MRI scanner (MagSense 360, Mindray, Shenzhen, China). The in-plane resolution varies in the range $0.6-1$ mm for both \textit{LF17} and \textit{LF19}, but slice thickness varies in the range $4$-$6$ mm and the gap in the range of $0.8$-$1.2$ mm for \textit{LF17} whereas slice thickness is $5$ mm and the gap in between slices is $1$ mm for \textit{LF19}. 

As a reference subset for low-field MRI simulations, we chose $28$ of the $39$ patients from \textit{LF17} where the MRIs are normal appearing according to a local radiologist. The remaining $11$ pathological brain scans were used for visualisation of lesions. Patients in \textit{LF19} were recruited and consented specifically for our study. Each underwent the standard local 0.36T protocol in Ibadan and travelled to a private clinic in Lagos, Nigeria, for high-field imaging. There, additional images were acquired on a 1.5T MRI scanner (Brivo MR355, GE Healthcare, Waukesha, WI, United States), including axial T1w (TE/TR = $17/720$ ms),  T2w (TE/TR = $102/6000$ ms) and FLAIR (TE/TR/TI = $120/6700/1678$ ms) images with an in-plane resolution of $0.47\times 0.47$ mm$^2$, slice thickness $5$ mm and gap between slices $1$ mm. A 3D fast spoiled gradient-echo T1w image (TE/TR = $5/13.5$ ms) with an in-plane resolution of $0.4\times 0.4$ mm$^2$and a slice thickness of $1.2$ mm was also acquired. 

An additional data cleaning step was required for processing the raw low-field images. Significant intensity differences arise between adjacent slices in some of the clinical low-field T1w scans, likely due to the cross-talk artifact. To correct this artifact, we selected a reference central slice with no apparent signal drop in each case and adjusted the intensity of the remaining slices by histogram matching in Matlab (The Mathworks, Natick, MA, United States). We also applied bias field correction to all low-field images using the N4 algorithm in Advanced Normalization Tools (ANTs, University of Pennsylvania, PA, USA).


\subsection{Implementation details}

We randomly selected training subjects ($60$ by default but more in some experiments where specified) from the HCP T1w data set and employed their images as the set of high-field images $Y$ from which to synthesise low-field images using Algorithm~\ref{alg:1}. The random selection excludes $100$ subjects who we reserve as an unseen test set. Low-field data from the $28$ historic normal-appearing MRIs in \textit{LF17}, provided default parameters for the distribution $P$ of GM and WM SNRs for low-field simulation by fitting a 2D Gaussian distribution to the collection of $28$ mean 0.36T GM and WM SNRs. We refer to this default distribution as $P_0$ and it has mean SNRs of $[64.50, 54.14]$ and covariance matrix $[[78.47, 71.50], [71.50, 73.91]]$. We used similar procedures for T2w and FLAIR images randomly sampled from the HCP T2w data set and the LEMON data set. The mean and the covariance matrix of SNRs are $[35.20, 48.46]$ and $[[84.15, 104.89], [104.89, 138.70]]$ for T2w, and $[35.99, 40.92]$ and $[[100.99, 74.34], [74.34, 129.56]]$ for FLAIR.

Our IQT program was implemented in Python using Tensorflow 2.0~\citep{tensorflow2016-whitepaper}. Processing used a GPU board with video random access memory (VRAM) larger than $12$ GB, e.g. Nvidia Tesla P100 (Nvidia, Santa Clara, CA, United States). Training the network used ADAM~\citep{kingma2014adam} as the optimizer with a starting learning rate of $10^{-3}$ and a decay of $10^{-6}$. We initialized the parameters with Glorot normal initializer~\citep{glorot2010understanding}. The batchsize was $32$. We monitored potential overfitting with a validation set splitting the full training set into $80\%$ and $20\%$ subsets of subjects for training and validation. All the experiments ran $100$ epochs to ensure the convergence of loss on the validation set.

\subsection{Anisotropic U-Net outperforms other architectures}\label{sec:5.1}
We start by demonstrating the benefits of our proposed anisotropic U-Net (ANISO U-Net) over standard architectures within a simple constrained task using simulated low-field images as the test set and with fixed SNR (i.e. the distribution $P$ is a Dirac's delta function) in both training and test sets. We fixed $SNR_X^{WM} = 64.50$ and $SNR_X^{GM} = 54.14$, which are the mean values of $P_0$. We evaluated performance for three upsampling factors $r=2,4,8$; the ratio of slice thickness and gap in the simulated low-field images was set to be $3:1$, which best reflects the real clinical low-field data. Specifically, the slice thickness/gap is $1.05$ mm/$0.35$ mm for $r=2$, $2.1$ mm/$0.7$ mm for $r=4$, and $4.2$ mm/$1.4$ mm for $r=8$.

We compare ANISO U-Net with standard 3D cubic B-spline interpolation, variants of both the Low-Rank Total-Variation (LRTV) method~\citep{Shi2015} and the 3D efficient sub-pixel convolutional neural network (ESPCN)~\citep{Shi2016,Tanno2021} for one-directional super resolution, and several existing U-Net baselines equivalent to switching off or substituting the bottleneck block and the residual block in ANISO U-Net. One is a 3D isotropic U-Net (ISO U-Net)~\citep{Cicek2016}, for which the input is isotropically interpolated using cubic B-splines. The other is 3D Super-Resolution U-Net (3D SR U-Net)~\citep{Heinrich2017} which up-samples each level output on the contraction path before concatenation. All the architectures of the U-Net variants have $5$ levels and $2$ convolutional layers per level, the number of filters on the first level is $16$ and doubles at each subsequent level.

In terms of image metrics, we evaluate performance on the test set by calculating peak signal-to-noise ratio (PSNR) and structural similarity index (SSIM)~\citep{Wang2004} between the original high-field image and that estimated from the simulated low-field image for each technique. 

Table~\ref{tab:3.3} shows that ANISO U-Net achieved the best performance in terms of the average PSNR and SSIM. As expected, reconstruction degrades as the $r$ increases for all methods, but ANISO U-Net consistently achieves the best scores with all three metrics. In particular, it significantly (two-tailed Wilcoxon signed-rank test with $p<0.001$) outperformed the others in terms of any metric and any examined $r$.  

\begin{figure}[!t]
    \centering
    \includegraphics[width=0.48\textwidth]{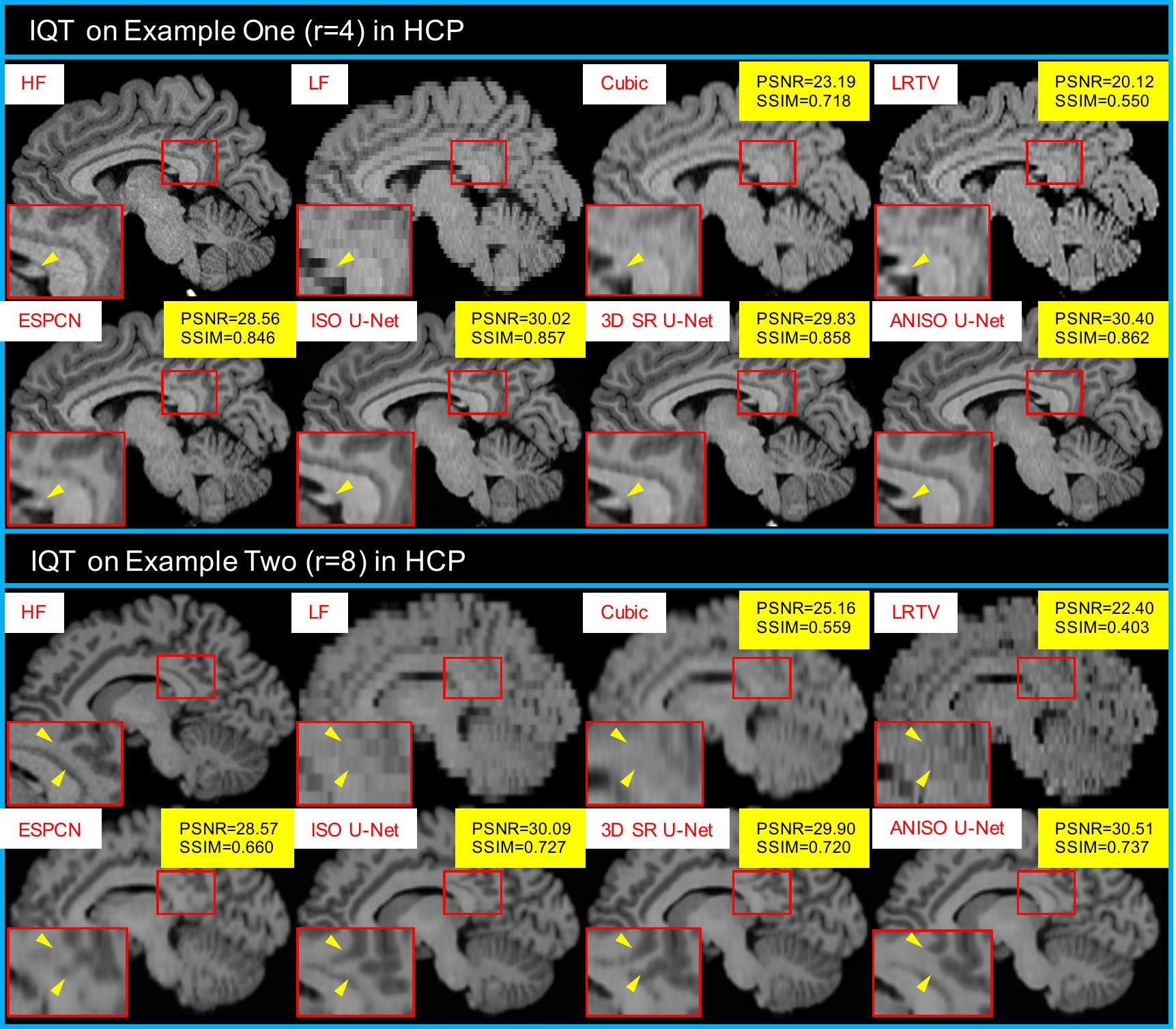}
    \caption{Sagittal visualisations of enhancing two example low-field images using various proposed implementations of IQT with fixed contrasts during training and two example upsampling factors $r=4$ and $r=8$. The figure compares, in two example T1w images, the high-field (HF) reference, the low-field (LF) input, output of cubic interpolation (Cubic), low-rank and total variation (LRTV) reconstruction and four deep-learning architectures described in the text.}
    \label{fig:3.3-sagittal}
\end{figure}

Figures~\ref{fig:3.3-sagittal} and~\ref{fig:3.3-coronal} show example U-Net reconstructions for both $r=4$ and $r=8$ in coronal and sagittal planes, respectively. Qualitatively we observe clear recovery of high resolution information and enhancement of contrast. Generally, the reconstructed images demonstrate the U-Net variants' ability to highlight features visible in the ground truth images that are obscured in the low-quality input. The zoomed patches in Figures~\ref{fig:3.3-sagittal} and~\ref{fig:3.3-coronal} highlight clear differences among algorithms and show examples where the ANISO U-Net approximates the ground truth most closely and with the least artefacts. Although the quantitative results in Table~\ref{tab:3.3} show only modest differences among the U-Net outputs, these global metrics are sometimes insensitive to localised differences that are clear qualitatively.

\begin{table}[!t]
\scriptsize
\centering
\caption{The performance of the listed models on upsampling factors $r=2,4,8$ in Section \ref{sec:5.1}. The mean and standard deviation (in bracket) of PSNR and SSIM were calculated over $100$ evaluation subjects. Bold font denotes the best mean. All metrics between any two comparing methods were statistically significant ($p$-value$<0.001$) by means of two-tailed Wilcoxon signed-rank test.
}
\begin{tabular}{|c|c|c|c|c|c|c|}
\hline
\multirow{2}{*}{Method} & \multicolumn{2}{c|}{$r=2$} & \multicolumn{2}{c|}{$r=4$} & \multicolumn{2}{c|}{$r=8$} \\\cline{2-7}
  & PSNR     & SSIM & PSNR     & SSIM & PSNR      & SSIM \\
\hline
\multirow{2}{*}{Cubic} & $26.10$ & $0.827$ & $24.92$ & $0.731$ & $23.56$ & $0.590$ \\
& $(2.51)$ & $(0.012)$ &$(2.51)$ &$(0.013)$ &$(2.51)$ &$(0.013)$ \\
\hline
\multirow{2}{*}{LRTV} & $22.49$ & $0.664$ & $21.55$ & $0.558$ & $20.46$ & $0.420$ \\
& $ (2.50) $ & $ (0.013) $ & $  ( 2.51)  $ & $  (0.011) $ & $ ( 2.52)$ & $ (0.009)$ \\
\hline
\multirow{2}{*}{ESPCN}  & $31.06$ & $0.877$ & $29.19$ & $0.808$ & $27.64$ & $0.698$ \\
&$(2.62)$ &$ (0.011)$ &$ (3.79)$ & $( 0.088)$ & $(2.50)$ & $(0.015)$ \\
\hline
ISO & $34.48$ & $0.929$ & $32.44$ & $0.871$ & $29.33$ & $0.768$ \\
U-Net &$(2.71)$ & $ (0.008)$ &$(2.57)$ &$ (0.014)$ &$(2.52)$ &$(0.014)$ \\
\hline
3D SR & $34.79$ & $0.932$ & $32.35$ & $0.873$ & $29.22$ & $0.762$ \\
U-Net &$(2.77)$ &$(0.008)$&$(2.63)$&$(0.014)$&$(2.51)$&$(0.018)$ \\
\hline
ANISO & $\textbf{35.11}$ & $\textbf{0.934}$ & $\textbf{32.66}$ & $\textbf{0.875}$  & $\textbf{29.69}$ & $\textbf{0.777}$ \\
U-Net &$(2.76)$&$(0.008)$&$(2.59)$&$(0.014)$&$(2.51)$&$(0.018)$ \\
\hline
\end{tabular}
\label{tab:3.3}
\end{table}

We next evaluate volume estimation for the above methods. We use a standard brain-segmentation tool, FastSurfer~\citep{Henschel2022}, to perform rapid subcortical segmentation of the human brain and report the relative volumetric error (RVE)~\citep{yousefi2012} in seven subcortical structures. We focus on subcortical structures, as first they are the most clinically relevant for our target application (childhood epilepsy), and second standard segmentation tools prove unreliable for cortical areas with our low-field images so a meaningful comparison is hard to make. The RVE is a measure of consistency of the estimated volume against the gold standard volume estimated from the original high field image; for the $i$th subcortical structure
\begin{equation}\label{eq:RVE}
\textrm{RVE}_i = \frac{2|V_i-V_{i}^*|}{V_i+V_{i}^*},
\end{equation}
where $V_i$ is the volume of the $i$th subcortical structure obtained from the low-quality or enhanced image and $V_{i}^*$ the gold-standard from the high-quality image.

\begin{figure*}[!t]
    \centering
	\includegraphics[width=1\textwidth]{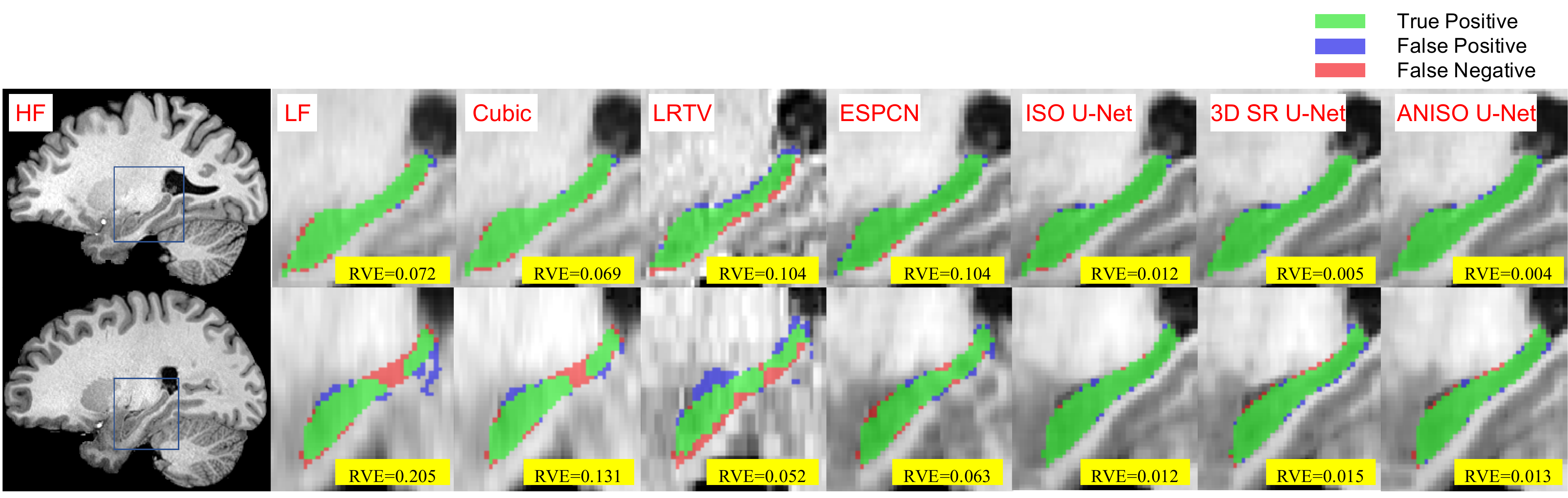}
    \caption{True positive, false positive, and false negative hippocampus maps as segmented by FastSurfer in low-field and enhanced images, compared to high field for two HCP examples for $r=4$ (first row) and $r=8$ (second row) in the sagittal plane. The set of methods compared is as described in the caption of Figure~\ref{fig:3.3-sagittal}. False positive and false negative regions indicate over- and under-estimation of hippocampal volumes.}
    \label{fig:3.3-2-aseg-hippo}
\end{figure*}

Table~\ref{tab:3.3-2} shows the mean and standard deviation RVE scores for 100 simulated HCP data sets with fixed contrast, obtained for the set of super-resolution and IQT methods we compare in other experiments. Our ANISO U-Net usually achieves the highest accuracy; in the few cases in which the ISO U-Net and the 3D SR U-Net provide higher accuracy, ANISO U-Net is a very close second or third. Figure~\ref{fig:3.3-2-aseg-hippo} gives an example volume segmentation highlighting true and false positive and false negative areas for the hippocampus for each candidate method compared to high field in two HCP data sets using $r=4$ and $r=8$ in the sagittal plane. The example is typical of most data sets.  The other U-Net methods often lead to small overestimation of the hippocampal volumes whereas other simpler methods have significantly larger both false positive and negative counts.

\begin{table*}[!t]
\scriptsize
\centering
\caption{The performance of the listed models on upsampling factors $r=2,4,8$ in Section \ref{sec:5.1}. The mean and standard deviation of RVE scores were computed over $100$ evaluation subjects with fixed contrast. Bold font denotes the best mean. 
}
\begin{tabular}{|c|c|c|c|c|c|c|c|c|}
\hline
Structure & $r$ & Input & Cubic	 & LRTV  & ESPCN & ISO U-Net & 3D SR U-Net & ANISO U-Net \\
\hline
Thalamus & \multirow{7}{*}{$2$} & $0.077\pm 0.013$&$0.082\pm 0.013$&$0.120\pm 0.017$&$0.019\pm 0.012$&$0.014\pm 0.007$&$\textbf{0.010}\pm 0.006$&$\textbf{0.010}\pm 0.005$ \\
Caudate& &$0.274\pm 0.083$&$	0.271\pm 0.082$&$0.280\pm 0.088$&$0.271\pm 0.088$&$0.030\pm 0.028$&$0.037\pm 0.030$&$\textbf{0.025}\pm 0.024$ \\
Putamen& &$0.229\pm 0.150$&$0.206\pm 0.144$&$0.103\pm 0.104$&$0.551\pm 0.205$&$0.040\pm 0.033$&$0.028\pm 0.020$&$\textbf{0.014}\pm 0.011$ \\
Palllidum& &$0.097 \pm 0.055$&$	0.093 \pm 0.056$&$	0.066 \pm 0.052$&$	0.141 \pm 0.069$&$	0.064 \pm 0.041$&$	0.047 \pm 0.031$&$	\textbf{0.027} \pm 0.020$ \\
Hippocampus& & $0.078 \pm 0.030$&$	0.077 \pm 0.030$&$	0.119 \pm 0.031$&$	0.023 \pm 0.019$&$	0.019 \pm 0.015$&$	0.016 \pm 0.013$&$	\textbf{0.015} \pm 0.011$ \\
Amygdala& &$0.122 \pm 0.031$&$	0.119 \pm 0.032$&$	0.160 \pm 0.044$&$	0.068 \pm 0.028$&$	0.035 \pm 0.021$&$	0.025 \pm 0.018$&$	\textbf{0.018} \pm 0.015$\\
Accumbens& &$0.351 \pm 0.058$&$	0.337 \pm 0.060$&$	0.238 \pm 0.050$&$	0.209 \pm 0.101$&$	0.029 \pm 0.023$&$	0.027 \pm 0.022$&$	\textbf{0.022} \pm 0.016$ \\
\hline
Thalamus & \multirow{7}{*}{$4$} &$0.074 \pm 0.018$&$	0.081 \pm 0.017$&$	0.107 \pm 0.018$&$	0.017 \pm 0.011$&$	0.028 \pm 0.010$&$	0.020 \pm 0.010$&$	\textbf{0.011} \pm 0.007$ \\
Caudate& &$0.311 \pm 0.094$&$	0.304 \pm 0.092$&$	0.322 \pm 0.096$&$	0.307 \pm 0.096$&$	\textbf{0.031} \pm 0.024$&$	0.042 \pm 0.033$&$	0.037 \pm 0.032 $\\
Putamen& &$0.345 \pm 0.185$&$	0.304 \pm 0.171$&$	0.165 \pm 0.139$&$	0.618 \pm 0.200$&$	\textbf{0.036} \pm 0.032$&$	0.046 \pm 0.029$&$	0.045 \pm 0.032$ \\
Palllidum& &$0.097 \pm 0.061$&$	0.099 \pm 0.060$&$	0.079 \pm 0.059$&$	0.129 \pm 0.075$&$	0.040 \pm 0.034$&$	\textbf{0.035} \pm 0.032$&$	0.067 \pm 0.043$ \\
Hippocampus& &$0.075 \pm 0.047$&$	0.081 \pm 0.045$&$	0.111 \pm 0.042$&$	0.035 \pm 0.028$&$	0.022 \pm 0.016$&$	0.017 \pm 0.016$&$	\textbf{0.016} \pm 0.013$ \\
Amygdala& &$0.129 \pm 0.039$&$	0.133 \pm 0.038$&$	0.137 \pm 0.043$&$	0.063 \pm 0.038$&$	0.033 \pm 0.022$&$	0.028 \pm 0.022$&$	\textbf{0.026} \pm 0.020$ \\
Accumbens& &$0.404 \pm 0.067$&$	0.392 \pm 0.066$&$	0.291 \pm 0.065$&$	0.274 \pm 0.112$&$	0.040 \pm 0.034$&$	\textbf{0.028} \pm 0.020$&$	0.050 \pm 0.031$ \\
\hline
Thalamus & \multirow{7}{*}{$8$} & $0.049 \pm 0.027$&$	0.054 \pm 0.031$&$	0.073 \pm 0.032$&$	0.034 \pm 0.028$&$	0.041 \pm 0.018$&$	\textbf{0.022} \pm 0.016$&$	0.024 \pm 0.015$ \\
Caudate& &$0.398 \pm 0.145$&$	0.337 \pm 0.110$&$	0.385 \pm 0.115$&$	0.318 \pm 0.104$&$	0.049 \pm 0.042$&$	0.070 \pm 0.049$&$	\textbf{0.046} \pm 0.041$ \\
Putamen& & $0.664 \pm 0.285$&$	0.484 \pm 0.223$&$	0.312 \pm 0.194$&$	0.708 \pm 0.227$&$	0.109 \pm 0.052$&$	0.069 \pm 0.056$&$	\textbf{0.064} \pm 0.038$ \\
Palllidum& &$0.129 \pm 0.120$&$	0.087 \pm 0.072$&$	0.081 \pm 0.059$&$	0.101 \pm 0.061$&$	0.089 \pm 0.054$&$	0.066 \pm 0.055$&$	\textbf{0.053} \pm 0.049$ \\
Hippocampus& &$0.290 \pm 0.170$&$	0.151 \pm 0.117$&$	0.091 \pm 0.087$&$	0.058 \pm 0.049$&$	\textbf{0.023} \pm 0.019$&$	0.029 \pm 0.021$&$	0.024 \pm 0.017$ \\
Amygdala& &$0.216 \pm 0.090$&$	0.154 \pm 0.068$&$	0.100 \pm 0.061$&$	0.107 \pm 0.065$&$	\textbf{0.027} \pm 0.022$&$	0.034 \pm 0.027$&$	0.028 \pm 0.024$ \\
Accumbens& &$0.603 \pm 0.167$&$	0.443 \pm 0.121$&$	0.360 \pm 0.093$&$	0.580 \pm 0.180$&$	0.057 \pm 0.041$&$	0.087 \pm 0.062$&$	\textbf{0.041} \pm 0.033$ \\
\hline
\end{tabular}
\label{tab:3.3-2}
\end{table*}

\subsection{Hyperparameters}\label{sec:5.2}

In this section we demonstrate the influence of key hyperparameters on the performance of ANISO U-Net within the same simple simulation experiment as the previous section. We show results only for $r=4$ and T1w images, but similar trends arise for other $r$ and contrasts such as T2w and FLAIR. For all experiments, the default set of parameters is as in the previous section: training/test sets derived from $60$/$100$ subjects respectively, patch size of $32$ with the extraction step of $16\times 16\times 4$, $5$ levels in the ANISO U-Net, and $16$ filters on the first level. Each experiment varies one hyper-parameter out of the following: number of subjects, patch size, extraction step, number of levels, and number of filters, with all others fixed to default values. The aim is to provide users of our algorithm a sense of the data, training, and capacity requirements for the algorithm.

Tables~\ref{tab:5.2-1}-\ref{tab:5.2-4} broadly show that the performance improves as the amount of training data (number of subjects and dense patch extraction) increases, as expected, and as the model capacity (specifically number of filters and levels) of ANISO U-Net increases. However, we observe diminishing returns as we approach high training set size and model capacity and conclude that our chosen reference is a good balance of performance against cost of training time and memory requirements. Table~\ref{tab:5.2-5} shows that patch size of $32$ outperforms both $16$ and $48$ that indicating it is a good operating point trading off structural information content with ability to learn and generalise from a finite training set. 

\begin{table}[!t]
\footnotesize
\centering
\caption{Quantitative performance of ANISO U-Net v.s. number of training subjects with $r=4$. The mean and standard deviation of PSNR and SSIM are calculated over $100$ evaluation subjects. Bold font denotes the best mean.}
\begin{tabular}{|c|c|c|}
\hline
\#Subjects          & PSNR (dB)       & SSIM  \\
\hline
$30$ & $30.96 \pm2.79 $ & $0.843 \pm0.015$   \\
$60$ & $32.63 \pm3.00 $ & $0.855 \pm0.016 $   \\
$120$ & $32.63 \pm2.94 $ & $0.870 \pm0.016 $   \\
$240$ & $32.89 \pm2.64 $ & $0.880 \pm0.015 $   \\
$480$ & $33.17 \pm2.65 $ & $0.882 \pm0.015 $   \\
$960$ & $\textbf{33.42} \pm2.67 $ & $\textbf{0.886} \pm0.015 $   \\
\hline
\end{tabular}
\label{tab:5.2-1}
\end{table}

\begin{table}[!t]
\footnotesize
\centering
\caption{Quantitative performance v.s. number of filters of ANISO U-Net with $r=4$. The mean and standard deviation of PSNR and SSIM are calculated over $100$ evaluation subjects. Bold font denotes the best mean.}
\begin{tabular}{|c|c|c|}
\hline
\#Filters          & PSNR (dB)      & SSIM \\
\hline
$16$ & $32.88\pm3.01$ & $0.862\pm0.016$   \\
$32$ & $33.06\pm3.03$ & $0.865\pm0.016$   \\
$64$ & $33.05\pm3.02$ & $0.869\pm0.016$   \\
$128$& $\textbf{33.53}\pm3.11$ & $\textbf{0.877}\pm0.016$   \\
\hline
\end{tabular}
\label{tab:5.2-2}
\end{table}

\begin{table}[!t]
\footnotesize
\centering
\caption{Quantitative performance v.s. number of levels of ANISO U-Net with $r=4$. The mean and standard deviation of PSNR and SSIM are calculated over $100$ evaluation subjects. Bold font denotes the best mean.}
\begin{tabular}{|c|c|c|}
\hline
\#Levels & PSNR (dB)      & SSIM \\
\hline
$3$ & $32.55\pm2.99$ & $0.858\pm0.015$   \\
$4$ & $32.66\pm2.99$ & $0.858\pm0.016$   \\
$5$ & $\textbf{32.75}\pm3.00$ & $\textbf{0.860}\pm0.016$   \\
\hline
\end{tabular}
\label{tab:5.2-3}
\end{table}

\begin{table}[!t]
\footnotesize
\centering
\caption{Quantitative performance of ANISO U-Net v.s. the patch-extraction step with $r=4$. The mean and standard deviation of PSNR and SSIM are calculated over $100$ evaluation subjects. Bold font denotes the best mean.}
\begin{tabular}{|c|c|c|c|}
\hline
Extraction Step & PSNR (dB)      & SSIM \\
\hline
$(8,8,2)$ & $32.52\pm2.64$ & $\textbf{0.874}\pm0.015$   \\
$(16,16,4)$ & $\textbf{32.86}\pm3.08$ & $0.863\pm0.017$   \\
$(32,32,8)$ & $30.39\pm2.53$ & $0.824\pm0.014$   \\
\hline
\end{tabular}
\label{tab:5.2-4}
\end{table}

\begin{table}[!t]
\footnotesize
\centering
\caption{The performance of ANISO U-Net v.s. the output patch size with $r=4$. The mean and standard deviation of PSNR and SSIM are calculated over $100$ evaluation subjects. Bold font denotes the best mean.}
\begin{tabular}{|c|c|c|}
\hline
Patch Size  & PSNR (dB)      & SSIM \\
\hline
$16$ & $31.58\pm2.57$ & $0.847\pm0.015$   \\
$32$ & $\textbf{32.42}\pm2.61$ & $\textbf{0.869}\pm0.015$   \\
$48$ & $31.70\pm2.57$ & $0.860\pm0.014$   \\
\hline
\end{tabular}
\label{tab:5.2-5}
\end{table}

\subsection{Contrast variation and normalisation during training}\label{sec:results_training_contrast_norm}
\begin{figure*}[!t]
     \centering
     \includegraphics[width=1\textwidth]{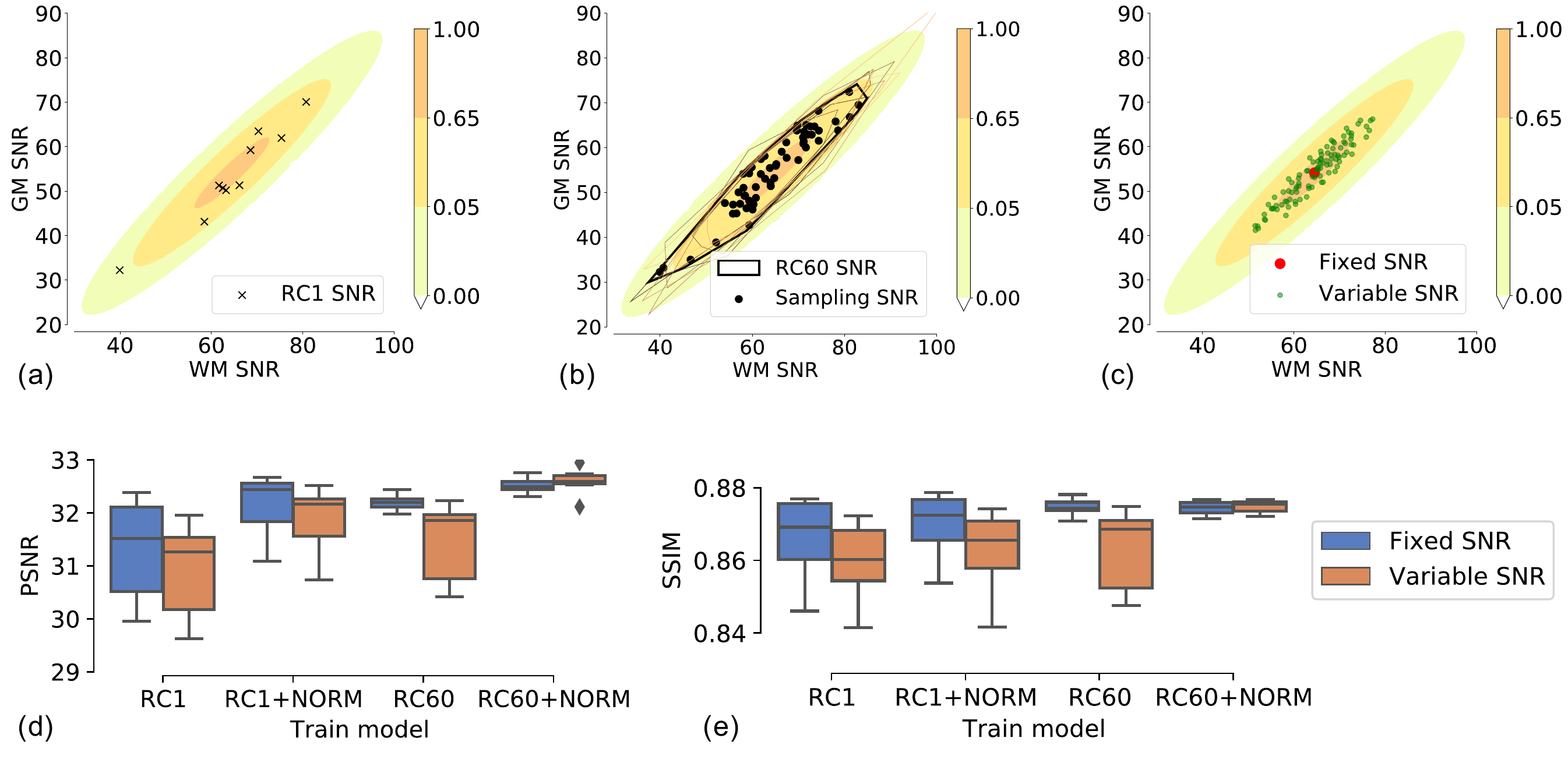}
       \caption{Evaluation of variable contrast and normalisation in the proposed IQT framework with $r=4$ for the experiments described in Section~\ref{sec:results_training_contrast_norm}. (a) Sampled contrast using for training RC1, (b) for RC60, and (c) test sampled contrasts for the experiments. (d) PSNR and (e) SSIM were averaged over the same $100$ evaluation subjects for the test cases of Fixed SNR (contrast does not vary among test images) and Variable SNR (contrast in test images is sampled from $P_0$). The mean and standard deviation of PSNR and SSIM are calculated over $10$ IQT models that were respectively trained with different random training sets, i.e. RC1 and RC60, and with/without NORM.}
     \label{fig:sec3.5-train-test-contrast}
 \end{figure*}

This section studies the impact of introducing variable contrast into the training data as well as the trainable histogram normalisation step at the training phase. We focus again on simulation experiments where ground truth is known and use $r=4$ to demonstrate. All contrasts were randomly drawn from the default distribution in Section 3.1. Two types of training sets were simulated, one with one random contrast (RC1) common to all $60$ subjects and the other with $60$ random contrasts (RC60) on the same $60$ subjects (each subject generates an image with a unique contrast drawn from $P_0$). We repeated the whole experiment $10$ times, each time with a different random seed generating a different set of training contrasts, as Figure~\ref{fig:sec3.5-train-test-contrast} (a-c) illustrates. We evaluated performance on two test sets: one with the fixed contrast of the mean SNRs $[64.50,54.14]$ of the reference Gaussian distribution for all $100$ subjects, denoted Fixed SNR, and the other with variable contrasts sampled from $P_0$ for the same $100$ subjects, denoted Variable SNR. 
 
\begin{figure}[!t]
    \centering
    \includegraphics[width=0.48\textwidth]{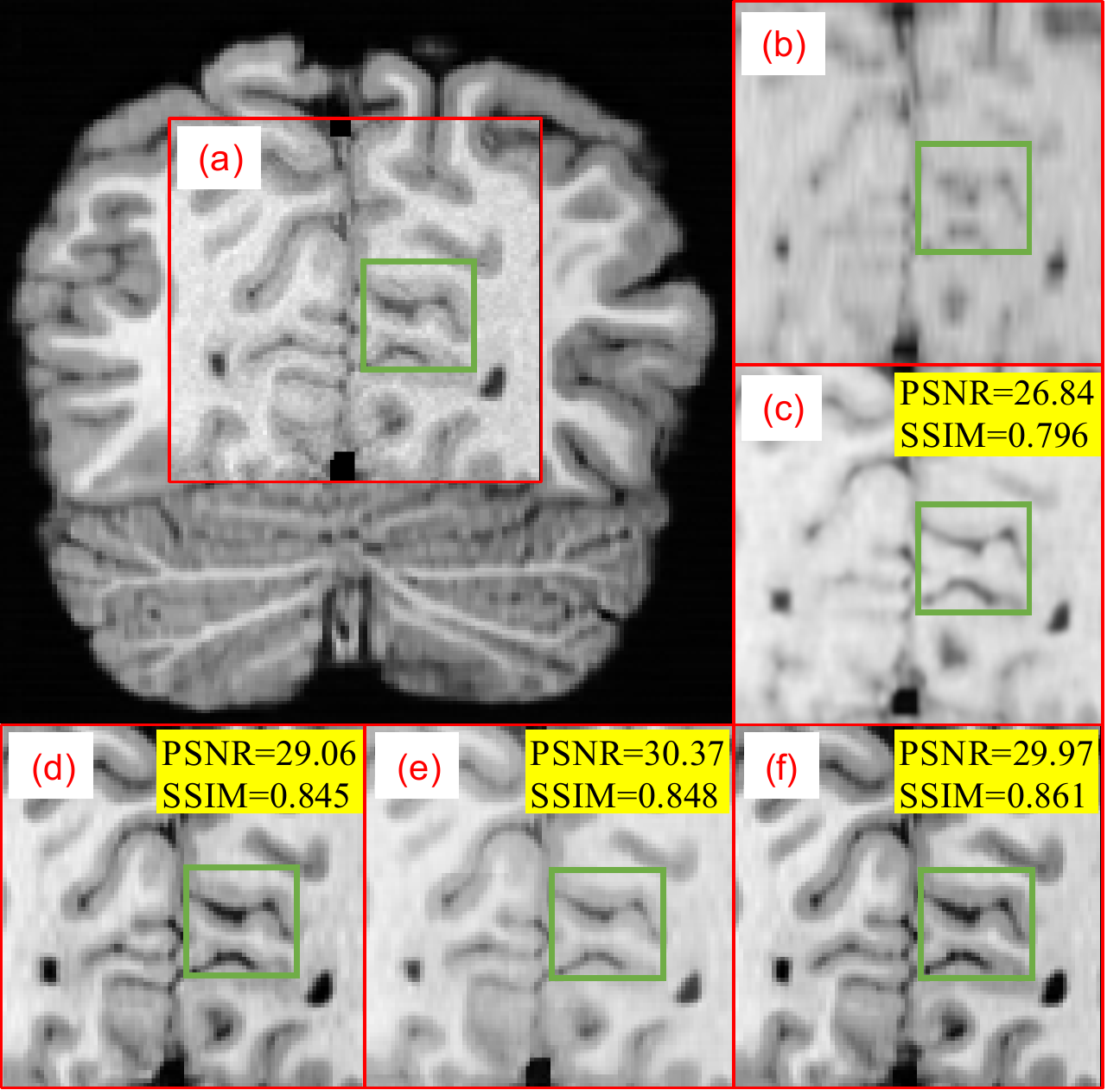}
    \caption{Visualisation and comparison of random-contrast IQT methods in the coronal plane of an example HCP data set. The top left panel shows the region of interest (ROI) of (a) a highlighted high-field image in the red box. The corresponding ROIs of (b) the low-field image patch, RC1 (c) without and (d) with NORM, RC60 (e) without and (f) with NORM are shown in the rest panels. For (c) and (e), i.e. without normalisation, the outputs show less good contrast (visually assessed). Including the NORM step in (d) and (f) retains the visual GM-WM contrast much more consistently.}
    \label{fig:3.5-vis-cor}
\end{figure}

\begin{figure}[!t]
     \scriptsize
     \centering
\includegraphics[width=0.48\textwidth]{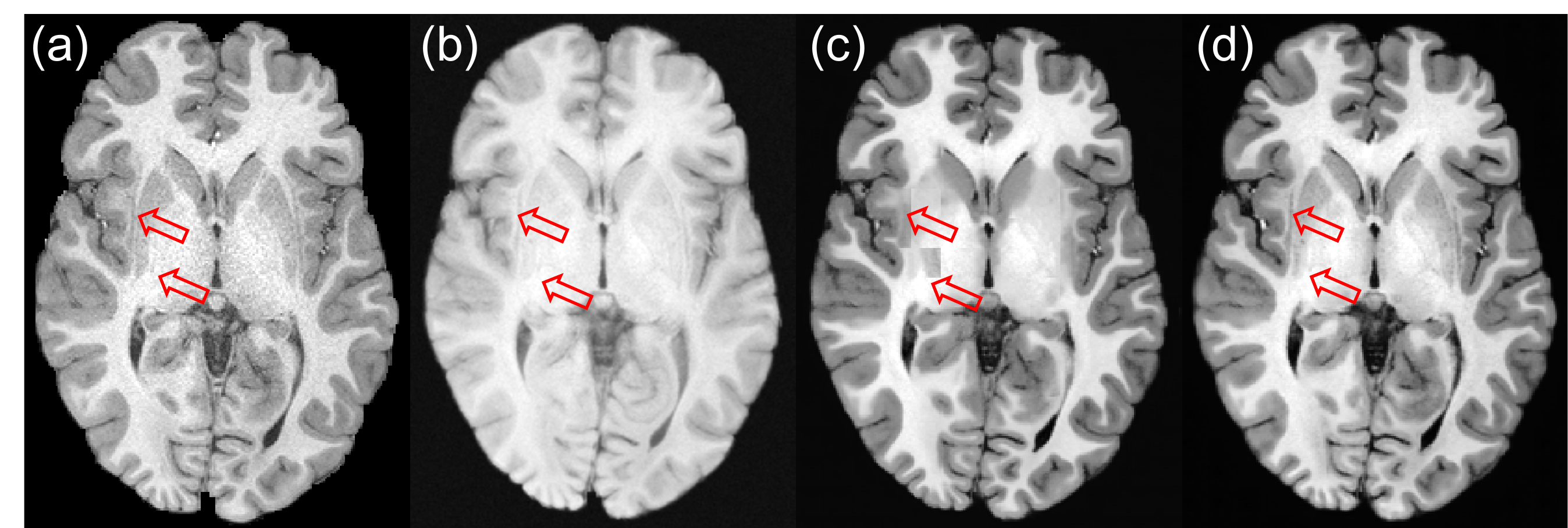}
     \caption{Axial visualisation of (a) the high-field, (b) the low-field, and (c) the RC1 IQT and (b) the RC60 IQT enhanced images both with NORM. Red arrows highlight corresponding spatial positions where tiling artefacts appear in RC1 IQT with NORM.}
     \label{fig:3.5-ax-tiling}
 \end{figure}

Figure~\ref{fig:sec3.5-train-test-contrast} compares the ability to recover high-field T1w reference images of models trained using each training set, and with and without the histogram normalisation step, which we refer to as NORM. We observed that statistically, the $10$ IQT models respectively trained on $10$ RC60 data sets had better average PSNR and SSIM scores than those respectively trained on $10$ RC1 data sets for both test cases in Figures~\ref{fig:sec3.5-train-test-contrast} (d-e). The difference was statistically significant when testing on the Variable-SNR data set and including the NORM step, which was examined by means of Wilcoxon sign rank test. The boxplots show that excluding the NORM step often leads to poor performance. Accordingly, incorporating the NORM step produces higher average performance with smaller variance over the 10 trials. Figures~\ref{fig:3.5-vis-cor} and~\ref{fig:3.5-vis-sag} visualise example outputs for the random-contrast IQT methods that were tested on the Fixed-SNR data and the Variable-SNR data, respectively. Generally, the main difference we observe among the methods is in contrast enhancement rather than resolution/sharpness. The use of the random-contrast IQT methods with the NORM step more clearly visualised the contrast difference and the sharpness of WM, GM and CSF even in the zoomed region than those without NORM. Moreover, in this experiment, although the NORM IQTs respectively with RC1 and with RC60 can show similar levels of image contrast, IQT with RC60 visually behaved better with less tiling artefacts than IQT with RC1; see Figure~\ref{fig:3.5-ax-tiling}.

\subsection{Radiological assessment}\label{sec:radioleval}

This section shows results and evaluations of radiological assessment on clinical image data from UCH Ibadan. We provide visual examples to illustrate the enhancement in typical normal and pathological cases and evaluate visual enhancement quantitatively in the normal brains via rating-scale questions to clinical radiologists. The system uses ANISO U-Net trained with the HCP and the LEMON data from $60$ subjects each with random contrasts for all three MR contrasts, respectively, and the default hyperparameters listed in Section~\ref{sec:5.2}. 

\begin{figure*}[!t]
    \centering
    \includegraphics[width=0.8\textwidth]{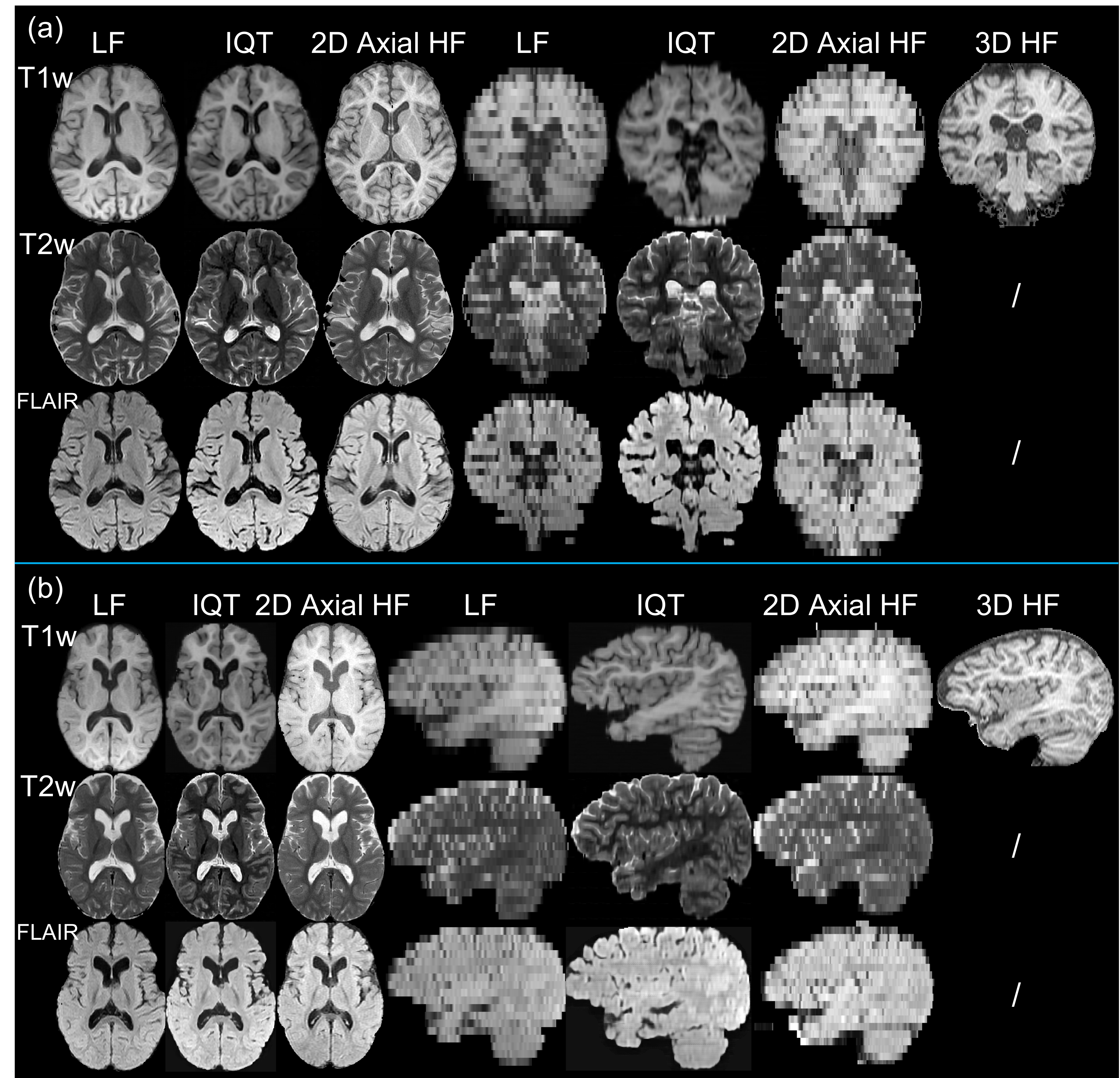}
    \caption{Two examples, (a) and (b), of IQT-enhanced T1w, T2w and FLAIR images with $r=8$, obtained from a 2D axial 0.36T low-field (LF) normal-appearing brain scan in the \textit{LF19} data set. The 2D axial and the 3D 1.5T high-field (HF) images are provided as high-quality references.}
    \label{fig:iqt-normal-vis1}
\end{figure*}

Figure~\ref{fig:iqt-normal-vis1} shows matched triplets of low-field, IQT-enhanced low-field, and high-field images for the three MR contrasts, i.e. T1w, T2w and FLAIR, from two example subjects in the \textit{LF19} data set. In general, the IQT enhanced images have greater contrast and show finer details than 0.36T references from the same subject. In particular, all three contrasts show substantial enhancement of detail in the slice direction, as we observe in the coronal and sagittal slices.  Contrast enhancement is particularly clear in the T1w images looking at the axial slices. 

Table~\ref{tab:radassess} shows that IQT enhances radiological assessment of the images. We asked two experienced radiologists to provide rating scores for each image for each of our $12$ neurological patients with matched low- and high-field images. Each radiologist was blinded to which image was which and provided scores between 1 (poor) and 4 (excellent) assessing a) WM-GM differentiation in the cerebrum, cerebellum and basal ganglia separately for each MR contrast; and b) visualisation of the hippocampus using coronal images and of the inferior frontal gyrus using sagittal images. All the images were originally acquired in axial orientation with 5-mm thick slices, so coronal and axial images are reformats of the original axial images for low-field and high-field images. The table shows that for GM-WM differentiation with T1w contrast, IQT significantly improves the radiologists evaluation compared to the original low-field ($p = 0.0124$ - two-tailed Wilcoxon signed-rank test) but remained worse than at high field ($p = 0.0026$ ); it also improves scores for T2w contrast significantly compared to low field ($p = 0.0104$) and scores even better than at high field (though not significantly). However, scores for the FLAIR contrast enhanced by IQT were not improved compared to low-field images (the overall score is slightly lower, but not significantly) and remained significantly worse than at high field ($p = 0.0105$). 

The visualisation of the hippocampus on coronal images and of the inferior frontal gyrus on sagittal images were evaluated using T1w, T2w and FLAIR images together. The average scores were 1.08/1.08 at 0.36T, 2.33/2.50 for IQT and 1.83/2.00 at 1.5T. These two structures are significantly better visualised on non-axial IQT images than at low field (p = 0.0156 for both), and even slightly better than at high field (not significantly), which arises because the IQT images have higher resolution through plane than the high-field images.

\begin{table*}[!t]
\footnotesize
\centering
\caption{The mean and the standard deviation of rating scores between 1 (poor) and 4 (excellent), on clinical LMIC low-field (LF) images, corresponding high-field (HF) images and the IQT-enhanced images. Two experienced radiologists separately reviewed clinical T1w, T2w, FLAIR scans from $12$ patients. They rated GM/WM differentiation in the cerebrum, cerebellum and basal ganglia in each contrast. They also rated the ability to delineate the hippocampus in the coronal plane (COR HIPPO) and inferior frontal gyrus in the sagittal plane (SAG IFG) for each contrast and each subject.}
\begin{tabular}{|c|c|c|c|c|c|c|}
\hline
\multirow{2}{*}{Type} & \multirow{2}{*}{Modal} & \multicolumn{3}{c|}{GM/WM Differentiation } & \multicolumn{2}{c|}{Visualisation } \\ \cline{3-7}
                       & & Cerebrum   & Cerebellum & Basal Ganglia & COR HIPPO & SAG IFG \\
\hline
\multirow{3}{*}{\shortstack{LF}} & T1w    & $2.08\pm 0.52$ & $1.67\pm 0.65$ & $1.67\pm 0.65$ & \multirow{3}{*}{$1.08\pm 0.29$} & \multirow{3}{*}{$1.08\pm 0.29$} \\
               					& T2w    & $2.67\pm 0.65$ & $2.75\pm 0.75$ & $2.83\pm 0.58$ & & \\
               					& FLAIR & $2.42\pm 0.67$ & $1.92\pm 0.67$ & $2.25\pm 0.62$ & & \\\hline
\multirow{3}{*}{\shortstack{IQT}}& T1w & $2.58\pm 0.79$ & $2.25\pm 0.87$ & $2.08\pm 1.00$ & \multirow{3}{*}{$2.33\pm 1.15$} &  \multirow{3}{*}{$2.50\pm 1.31$} \\
               					& T2w & $3.17\pm 0.72$ & $3.08\pm 0.79$ & $3.25\pm 0.75$ & & \\
               					& FLAIR & $2.42\pm 0.51$ & $1.83\pm 0.39$ & $2.17\pm 0.72$ & & \\\hline
\multirow{3}{*}{\shortstack{HF}} & T1w & $3.08\pm 0.79$ & $3.00\pm 0.85$ & $3.08\pm 1.08$ &  \multirow{3}{*}{$1.83\pm 0.39$} &  \multirow{3}{*}{$2.00\pm 0.60$} \\
               					& T2w & $2.92\pm 0.29$ & $3.25\pm 0.62$ & $3.17\pm 0.58$ & & \\
               					& FLAIR & $2.67\pm 0.78$ & $2.25\pm 0.75$ & $2.75\pm 0.62$ & & \\
\hline
\end{tabular}
\label{tab:radassess}
\end{table*}

\begin{figure*}[!t]
    \centering
    \includegraphics[width=0.9\textwidth]{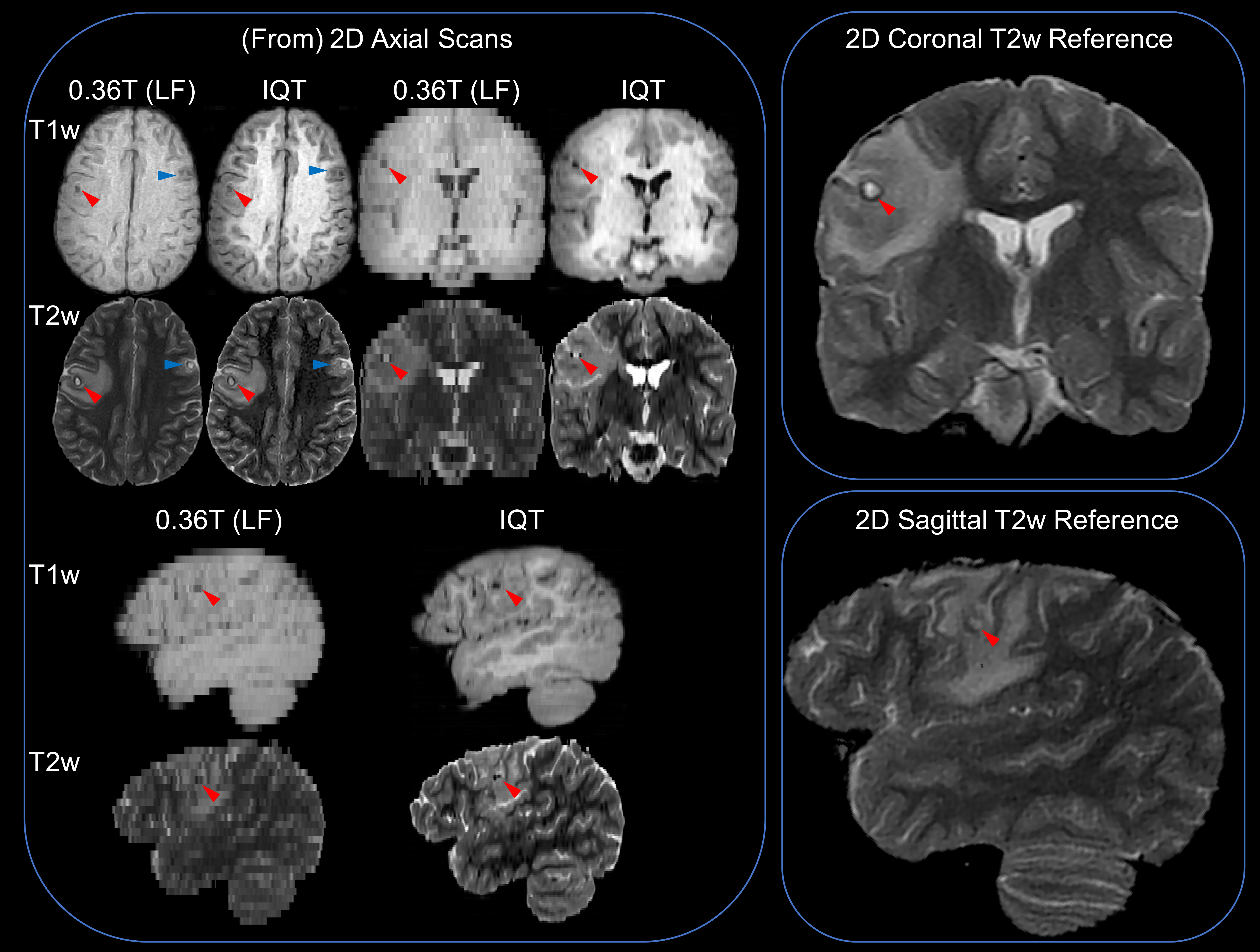}
    \caption{IQT-Enhanced T1w and T2w images with $r=4$ obtained from real low-field (LF) images of a pathological subject in the \textit{LF17} data set. Red and blue arrows point out two cystic lesions, with the diameter of $4.4$ mm and $5.5$ mm respectively, at the parietal lobes. 2D sagittal and coronal T2w references are provided to clearly visualise those two lesions.}
    \label{fig:iqt-lesion-n17-004}
\end{figure*}

Figures~\ref{fig:iqt-lesion-n17-004} and~\ref{fig:iqt-lesion-n19-012} demonstrate the ability of our IQT algorithm to enhance the conspicuity of clinically relevant lesions from the \textit{LF17} and the \textit{LF19} data sets, respectively. Figure~\ref{fig:iqt-lesion-n17-004} shows the impact of our IQT enhancement on clinical data from UCH Ibadan obtained from a 10-year-old epilepsy patient who has two cortical-subcortical parietal cystic lesions with surrounding edema. The lesions (red and blue arrows), respectively $4.4$ mm and $5.5$ mm in diameter, are weakly visible on low-field T1w images at the GM-WM junction of the parietal lobes; the edema is clear on the T2w images. The IQT approach improves the GM-WM contrast globally, and significantly enhances the resolution in the slice direction (i.e. in coronal and sagittal planes). The enhanced image strongly highlights the two lesions in this patient which are very subtle on the input T1w image. In this particular patient, the lesions were clearly visible on the coronal and the sagittal planes of the reference T2w images, which confirms that IQT highlights the lesions in the correct locations. Figure~\ref{fig:iqt-lesion-n19-012} illustrates the enhanced image quality of T1w, T2w and FLAIR images for a 12-year-old patient data showing middle cerebral artery ischemia. Our stochastic IQT enhancement greatly improves the resolution and visibility of the lesions. In this case, the original T1w low-field images show cross-talk and field inhomogeneity artifacts, but they were corrected during IQT pre-processing.

\subsection{Quantitative evaluation on clinical data}\label{sec:realquanteval}

Finally, we demonstrate volume estimation on the real-world data. With the same experimental setup as in Section~\ref{sec:5.1}, we used the RVE score to evaluate the segmentation accuracy on seven subcortical structures. Table~\ref{tab:results-LF19-segmentiation} shows RVE scores for the subcortical structures of low-field and IQT-enhanced images for T1w contrast calculated from five normal-appearing brain scans in the \textit{LF19} data set. We observed that the RVE scores of IQT in all the brain regions substantially improve compared to those of low-field, especially in Putamen and Hippocampus in terms of Reduction of RVE.

\begin{table}[!t]
\scriptsize
\centering
\caption{RVE scores for the seven subcortical structures of low-field and IQT-enhanced images for T1w contrast, calculated from five normal-appearing brain scans in the \textit{LF19} data set. High-field scans from the same subjects provide the gold standard. Reduction of RVE is calculated as the difference between the mean low-field and the mean IQT RVE scores divided by the IQT RVE score.
}
\begin{tabular}{|c|c|c|c|}
\hline
Structures & LF RVE   & IQT RVE 	 & Reduction of RVE  \\
\hline
Thalamus	&$1.788 \pm 0.209$&$	1.033 \pm 0.562 $ & $72.99\%$ \\
Caudate &$	1.792 \pm 0.188	$&$0.905 \pm 0.579 $ & $97.98\%$ \\
Putamen &$	1.790 \pm 0.163$&$	0.775 \pm 0.577 $ & $131.07\%$ \\
Palllidum &$1.727 \pm 0.243$&$	1.094 \pm 0.578$ & $57.82\%$ \\
Hippocampus &$	1.557 \pm 0.481$&$	0.657 \pm 0.366$ & $ 136.79\%$ \\
Amygdala	&$1.541 \pm 0.510$&$	0.879 \pm 0.568$ & $75.37\%$ \\
Accumbens &$	1.569 \pm 0.474	$&$1.031 \pm 0.507 $ & $52.22\% $ \\
\hline
\end{tabular}
\label{tab:results-LF19-segmentiation}
\end{table}

\section{Discussion}

In this work, we present a novel adaptation of IQT for application to standard structural MRI at low field. In particular, we estimate the contrast and resolution of high-field images (e.g. 1.5T or 3T) given clinical low-field images (e.g. 0.36T). The adaptation of IQT exploits two key innovations: (1) we introduce the ANISO U-Net, which naturally handles isotropic voxel sizes in super-resolution network architectures; (2) we devise a stochastic decimation simulator to capture variability in contrast in low-field images that arise in clinical practice and provide trained models with the robustness to generalise to such variations. We demonstrate with simulated data that the proposed method improves the robustness on unseen test data of variable SNR. Simulations further identify a viable working ``operating point’’ (combination of hyperparameters) with reasonably low data requirements and memory footprint but strong and robust performance. Finally, we demonstrate our algorithm on real-world clinical data from a LMIC hospital, showing encouraging qualitative improvements on both normal and pathological data sets and quantifying radiologists’ preference for the enhanced image for various standard brain-image analysis tasks. 

We have implemented a first viable prototype system for enhancement of clinical low-field MRI scans within the target environment of LMIC hospitals. This system demonstrates sufficient potential for supporting radiological evaluation and clinical trials of the method in practice. In the meantime, further developments will be investigated to produce a system with greater accuracy and robustness through better data sets, better evaluation and optimisation of operating points, better forward models, and better inverse models. We discuss the merits of the current implementation and these future developments in the rest of this section.

\subsection{General considerations and limitations}

Relying on a local, patch-based approach, our IQT algorithm does not alter the macroscopic structure of the image but enhances local resolution and contrast while retaining the gross morphology visible in the input image. For this reason, it cannot ameliorate image artifacts or distortions whereas more global image enhancement techniques offer such possibilities. On the other hand, this makes the algorithm unlikely to hallucinate erroneous macrostructure, to which global algorithms can be susceptible.

The patch-based approach also makes the trained IQT method less prone to overfitting, more memory-efficient, and more generalisable than whole-image methods, as small-scale features should be less variable than large-scale ones among individuals and between healthy and pathological brains. However, patching can lead to a loss of continuity in the resulting images. Here we use a simple patch-blending approach, which we find sufficiently effective given enough training data. Subtle tiling artifacts may still remain~\citep{Innamorati2020} and future work will consider more sophisticated stitching strategies.

As in most previous IQT applications, training data is generated from high-quality data sets crudely emulating the corresponding low-quality images. While this reduces the realism of the output, as noted in Section~\ref{sec:introduction}, the overall aim is to enhance low-field images rather than perfectly reproduce high-field images and even simple simulated data proves sufficient to support models that are effective for this purpose.  Moreover, simulated training data avoids the need for large data sets of matched high-/low-quality image pairs, which are hard to find or acquire, and avoids problems with misalignment that arise even when such data is available.

Performance of the current IQT implementation may reduce  as the test data departs from the training data. Therefore, using a model trained on images from one scanner or imaging sequence/protocol to enhance images from a different one will have lower performance than enhancing images from the same set-up. Histogram normalisation may mitigate such effects to some extent, but more advanced tools such as domain generalisation technology~\citep{liu2020shape} may further aid generalisability in the future.

\subsection{Better quality of data}

The ideal future scenario for IQT is to train and evaluate on larger and more diverse data sets than those used in the current experiments. Nevertheless, our preliminary results are compelling and sufficient to motivate acquisition of larger clinical data sets and evaluation of the impact of the IQT enhancement on clinical assessment and diagnostic decision. While the T1w and T2w high-field images from the HCP data set proves sufficiently extensive to train IQT effective models, the FLAIR data we use from the LEMON data set is of relatively lower quality compared to the HCP data set and the image contrast looks slightly different from our low-field and high-field acquisitions, consequently IQT-enhanced FLAIR images show sub-optimal performance levels compared to T1w and T2w contrasts.  Our on-going work is refining and packaging our software to substantially improve FLAIR results by means of better suited high-field FLAIR data sets.

Another option for future training data is to adopt real pairs of low-field and high-field data directly. At present, we prefer synthetic low-field data for IQT training for two practical reasons: First, sufficient quantity of paired data is difficult to obtain, because low-field scanners representative of LMIC settings are hard to find in HICs and acquiring data from high-field scanners is not routine in LMICs; we continue to acquire paired data in Nigeria, but the logistics of transportation of patients between sites is complex and expensive and we believe the limited data we do acquire is better used for validation purposes. Second, even if sufficient paired data were to become available, misalignment between pairs of images substantially complicates IQT model training. Alignment remains imperfect even after employing the best available image registration technology, particularly for low and high-field images with quite different appearance, and even small misalignments of training pairs introduces blurring in predicted high-field/high-quality images~\citep{Tax2020}. Recent advances~\citep{Kong2021} in image-to-image transformation learning attempt to mitigate misalignment via simultaneous learning of the transformation together with a misalignment map; while such ideas are promising, they are not well-established yet and we leave their development and evaluation for future work.

\subsection{Better implementation and evaluation}
The combination of settings and hyperparameters (operating point) that we identify for the working IQT system we run on clinical data in Section~\ref{sec:radioleval} was largely driven by practicality. We sought the simplest architecture and minimal data requirements that support a viable system.  Reducing the size of the training set from our choice of 60 high-field-low-field pairs (with patch sampling frequency fixed) can introduce tiling artefacts and false contrast into the reconstructed images, which we do not observe at the chosen operating point, although this would need verification on a larger test set before wider deployment. Increasing the size of the training set (as in Table~\ref{tab:5.2-1} for example) increases computational requirements for training rapidly and we wish to keep the system, both training and testing phases, viable for modest computational platforms such as those available in, or deployable to, LMIC settings. Nevertheless, prior to clinical trials, future work will need to re-evaluate on larger validation and evaluation sets the choice of hyperparameters, training set size, patch-sampling strategy, etc. to optimise choices for practical deployment in particular scenarios.

The current study was designed for deployment of IQT in LMICs. While model training remains intensive and needs substantial computational resource restricted mostly to HIC scenarios, model prediction at test time is achievable with modest computational resources. For example, an Intel Core i7 CPU (Santa Clara, CA, USA) with 16GB RAM is sufficient to run a trained IQT model and generate an output in around $68.22$ seconds per subject in our settings. Even in the absence of such resources, application of IQT remotely using HIC partner’s resources or cloud platforms is feasible. As the usage of IQT becomes more widespread, implementation within scanner-vendor platforms may become desirable with current on-board computational resources.

The global metrics (PSNR and SSIM) used for quantitative assessment reflect the performance only on synthetic images, since even small misalignment between images acquired on different platforms irrevocably disrupt such pixel-by-pixel comparisons~\citep{Tax2020}. Our qualitative assessment of how IQT improves radiological assessment and volumetric analysis through automated segmentation in Sections~\ref{sec:radioleval} and~\ref{sec:realquanteval} shows preliminary promise for clinical benefit using real-world data. However, evaluation on a large-scale data set of clinical images and verification of clinical significance from radiologists are essential for further translation. Furthermore, additional qualitative evaluation by radiologist ratings and, ultimately, demonstration of improved decision making is essential to confirm impact of the approach.

\subsection{Better modelling}\label{sec:better_modelling}

One key advance over previous IQT implementations is the stochastic nature of the forward model we use to generate synthetic low-field images for training. The advance is necessary to capture the variation of real-world clinical low-field images; failure to account for that variation can introduce bias in the trained model and diminish performance when enhancing real low-field images. However, our model remains a crude approximation capturing low-field WM-GM contrast via a simple 2D Gaussian distribution estimated from a small sample set of domain-specific data.  A conceptually appealing alternative is to employ a biophysical model that relates the MR signal at low and high fields to underlying tissue properties considering the precise scanner specification and sequence parameters used for each acquisition. In theory, such a model could underpin a more generalisable IQT mapping that accounts for the precise low-field MR sequence/scanner when estimating the corresponding high-field image. The literature includes various mathematical models that attempt to describe the variation of the MRI signal with the magnetic field strength~\citep{Marques2019,Brown2014}, but such models are simplistic and exclude or simplify many aspects of the imaging process that can have a strong influence particularly on low-field images. In practice they fail to capture the highly non-linear mapping from high to low field, which depends acutely on precise details of the imaging protocol; see for example the empirical study by~\citet{Wu2016} and the experiment by~\citet{Liu2021}. Other authors, e.g.~\citet{Xiang2018}, attempt to leverage adversarial training using unpaired training data but a) tests are limited to mappings between 3T and 7T where transformations are more linear and simpler to model and ignore resolution change and b) mitigation of anecdotal introduction of spurious artefacts~\citep{Cohen2018} by such approaches remains an area for future work. With these observations in mind, while we acknowledge limitations in our simple observational and data-set-specific forward model, we consider more sophisticated approaches an area for future work. 

Here we use a simple but effective backbone of convolutional neural networks for IQT. Nevertheless, the wider literature on super resolution reconstruction and modality transfer includes an increasingly diverse set of the advanced deep learning technologies that may further extend IQT to a broader set of scenarios, such as zero-shot super resolution (SR)~\citep{shocher2018zero}, progressive training for large scale SR~\citep{Ahn2018}, real-time quantised SR~\citep{Ignatov2021}, multi-frame fusion SR network~\citep{bhat2021deep}, bijective low resolution and high resolution mapping using hierarchical conditional flow~\citep{liang2021hierarchical} and Unsupervised SR with Cycle-in-Cycle structure~\citep{Yuan2018}. Enhanced deep neural network architectures, such as SR Generative Adversarial Network (SRGAN)~\citep{Ledig2017}, Enhanced Deep SR network (EDSR)~\citep{Lim2017}, Residual Channel Attention Network (RCAN)~\citep{Zhang2018Image}, and SR via repeated refinement (SR3) may offer potential benefits to performance. Alternative loss functions such as perceptual loss~\citep{johnson2016perceptual}, cycle-consistency loss~\citep{Ravi2019}, the informational noise-contrastive estimation (InfoNCE) loss~\citep{Wang2021cvpr} may facilitate better training models. In particular Generative Adversarial Network (GAN) architectures show spectacular achievements in generating realistic photographs of different styles and content~\citep{isola2017image, zhu2017unpaired, Zhu2017nips} and realistic modality translation in medical images~\citep{Wolterink2017, Maspero2018, Dar2019, Armanious2020}. 

Applications of such techniques to medical imaging must be evaluated with care to avoid artefacts that mislead radiologists e.g. as introducing lesions, or vice versa obscuring lesions or other pathologies not represented in the training set~\citep{Cohen2018}. The recent SynthSR techniques, such as~\citet{Iglesias2021,Iglesias2022}, fills in lesions intentionally to produce images that standard processing pipelines can handle, but are therefore not appropriate for clinical assessment. Our algorithm is relatively invulnerable to such problems, as the patch-based approach is local in nature and mostly learns the mapping of contrast rather than large-scale image structure. Nevertheless, recent examples such as~\citet{Ravi2022} highlight the potential of generative adversarial training in medical image synthesis. Transformer architectures also offer promise for super resolution, image reconstruction and generation, potentially boosting performance over CNNs by capturing long-distance structural relationships better. Extensions of such architectures from 2D to 3D remain preliminary and care is needed to mitigate the ill-posed nature of IQT and super-resolution particularly for large up-sampling factors~\citep{Yu2018}. The IQT technique is distinct from super-resolution in computer vision~\citep{Yang2019,Wang2021pami}, which is particularly challenging when upsampling factor is not less than $8$ times~\citep{Lai2019,Yang2020,lugmayr2021ntire}. 

Recent advances in IQT~\citep{Tanno2021,Finck2022} include simultaneous estimation of uncertainty and IQT enhancement. Although the current IQT implementation includes the potential for uncertainty estimation via the Masksembles layers mentioned in section~\ref{sec2.1}, we do not use uncertainty maps in our experiments and evaluations. Direct inclusion of IQT uncertainty mapping in~\citet{Tanno2021} is not straightforward here, because we work on non-quantitative structural images, where the overall appearance is more important than the quantitative voxel values, unlike diffusion MRI used in~\citet{Tanno2021}. Moreover, early-stage consultation with representative end-users (radiologists in the UK and Nigeria) for this project suggested we keep the output of our method as simple and immediately interpretable as possible, avoiding additional information that could confuse radiologists unfamiliar with artificial intelligence applications. Thus we leave full consideration of uncertainty quantification for future work. Nevertheless, for completeness, \ref{appendix:uq} shows preliminary results of uncertainty maps in Figures~\ref{fig:uq-hcp-ax-co} and~\ref{fig:uq-n17-ax-sa}, which show some promise by correlating well with error maps in simulation and highlighting pathological features that do not appear in our training data.

\section{Conclusion}

We introduce an IQT system that enhances low-field MRI and demonstrates efficacy in the LMIC clinical environment. However, future work is required to validate its potential impact on clinical decisions and enhance diagnostic accuracy for example in neurological conditions such as epilepsy, thereby making optimal clinical decisions, surgical planning and guided interventions difficult, where subtle lesions may be missed due to low image resolution and/or poor contrast. The IQT system shows potential to enhance subtle lesions in neurological conditions. IQT enhancement offers the potential to improve diagnostic confidence with low-field MRI systems in LMICs, as well as subsequent treatment planning and patient outcomes, as confirmed by neuroradiological assessment. This motivates future clinical trials and field studies to evaluate the clinical impact of IQT tools in LMICs and the development of interfaces that can exploit its benefits while mitigating potential mis-interpretation from processing artefacts. IQT systems similar to that we propose here offer great potential in the realisation and practical deployment of other low-field systems in particular portable MRI systems such as the Hyperfine Swoop~\footnote{https://hyperfine.io/} and even lower-field systems~\citep{Liu2021,Iglesias2022,VanSpeybroeck2021}.

\section*{CRediT authorship contribution statement}
HL: Methodology, Conceptualization, Data curation, Formal Analysis, Software, Investigation, Visualization, Validation, Writing – original draft, Writing – review \& editing. MF: Methodology, Data curation, Formal Analysis, Software, Visualization, Writing – review \& editing. FD: Validation, Visualization. GO: Resources, Validation, Writing – review \& editing. RT: Methodology. SBB: Methodology, Writing – review \& editing. LR: Methodology, Validation, Writing – review \& editing. BJB: Resources. DWC: Resources, Funding acquisition. IL: Resources. JHC: Resources, Funding acquisition. DFR: Funding acquisition, Writing – review \& editing. DCA: Project administration, Funding acquisition, Conceptualization, Supervision, Methodology, Writing – review \& editing.

\section*{Declaration of competing interest}
The authors declare that they have no known competing financial interests or personal relationships that could have appeared to influence the work reported in this paper.

\section*{Acknowledgements}
This work was supported by EPSRC grants (EP/R014019/1, EP/R006032/1 and EP/M020533/1), the NIHR UCLH Biomedical Research Centre, and the NIHR Biomedical Research Centre at Great Ormond Street Hospital. 3T T1w and T2w images were provided in part by the Human Connectome Project, WU-Minn Consortium (Principal Investigators: David Van Essen and Kamil Ugurbil; 1U54MH091657) funded by the 16 NIH Institutes and Centers that support the NIH Blueprint for Neuroscience Research; and by the McDonnell Center for Systems Neuroscience at Washington University. High-field FLAIR images were used in part from the Leipzig Study for Mind-Body-Emotion Interactions (LEMON) data set provided by the Mind-Body-Emotion group at the Max Planck Institute for Human Cognitive and Brain Sciences. For the purpose of open access, the author has applied a Creative Commons Attribution (CC BY) licence to any Author Accepted Manuscript version arising.

 \appendix
\section{Uncertainty mapping and preliminary results}\label{appendix:uq}
Generation of IQT uncertainty maps was a key innovation in the recent Bayesian IQT algorithm~\citep{Tanno2021}. However, the approach does not adapt naturally to the standard clinical scenario we consider here where images are non-quantitative: contrast is more important than absolute pixel value. We consider instead a recent alternative, the Masksembles layer~\citep{Durasov2021}, which accelerates (compared to~\citet{Tanno2021}) the estimation of ensemble-based model uncertainty and has shown efficacy in other problems. We include the approach as an optional component of our IQT architecture. The core idea of Masksembles is to use a mini-batch of weakly correlated random binary masks to drop out network weights during both training and test phases. We plug the Masksemble Layers directly after the convolutional layers within the backbone of ANISO U-Net. Masksembles has a $m$-by-$c$ binary matrix representing $m$ masks with $c$ dimensions.  The batch size is required to be divisible by $m$ so that all the masks can feed in one or several mini-batches of data in parallel. $c$ must be equal to number of channels of the last layer. The scale parameter $s$ controls the overlap of generated masks, usually less than $6$.

As a preliminary evaluation of the idea, we estimate uncertainty of IQT using the Anisotropic U-Net with the Masksembles layers of $16$ masks. We use the same data set as RC60 with NORM from Section~\ref{sec:results_training_contrast_norm} for training and evaluation. We input $16$ copies of a image patch as a batch simultaneously into the network and compute the Variance (Var) over the $16$ outputs as the measure of uncertainty in each voxel. 
Figure~\ref{fig:uq-hcp-ax-co} visualises an example set of synthetic low-field (LF), high-field (HF), IQT, error map (EM) against known ground truth (from the simulation), and uncertainty map (UM) in both axial and coronal planes. In a similar way to results in~\citet{Tanno2021} showing efficacy of uncertainty estimates, we observe that the UM is generally high in regions where the EM has high variance (noisy speckled patterns). This shows, as we would hope, that uncertainty estimates are high where the error distribution accommodates high values. Figure~\ref{fig:uq-n17-ax-sa} shows uncertainty maps on a real-world data set from \textit{LF17} (the same data set used in Figure~\ref{fig:iqt-lesion-n17-004}). The maps highlight some lesions, unseen in the training data, but not all and generally highlight white matter regions. This underlines the challenges with uncertainty estimation in non-quantitative images and prompts further work on the meaning and implementation of UM in such scenarios.

\section{Supplementary data}
\begin{figure}[!t]
    \centering
    \includegraphics[width=0.48\textwidth]{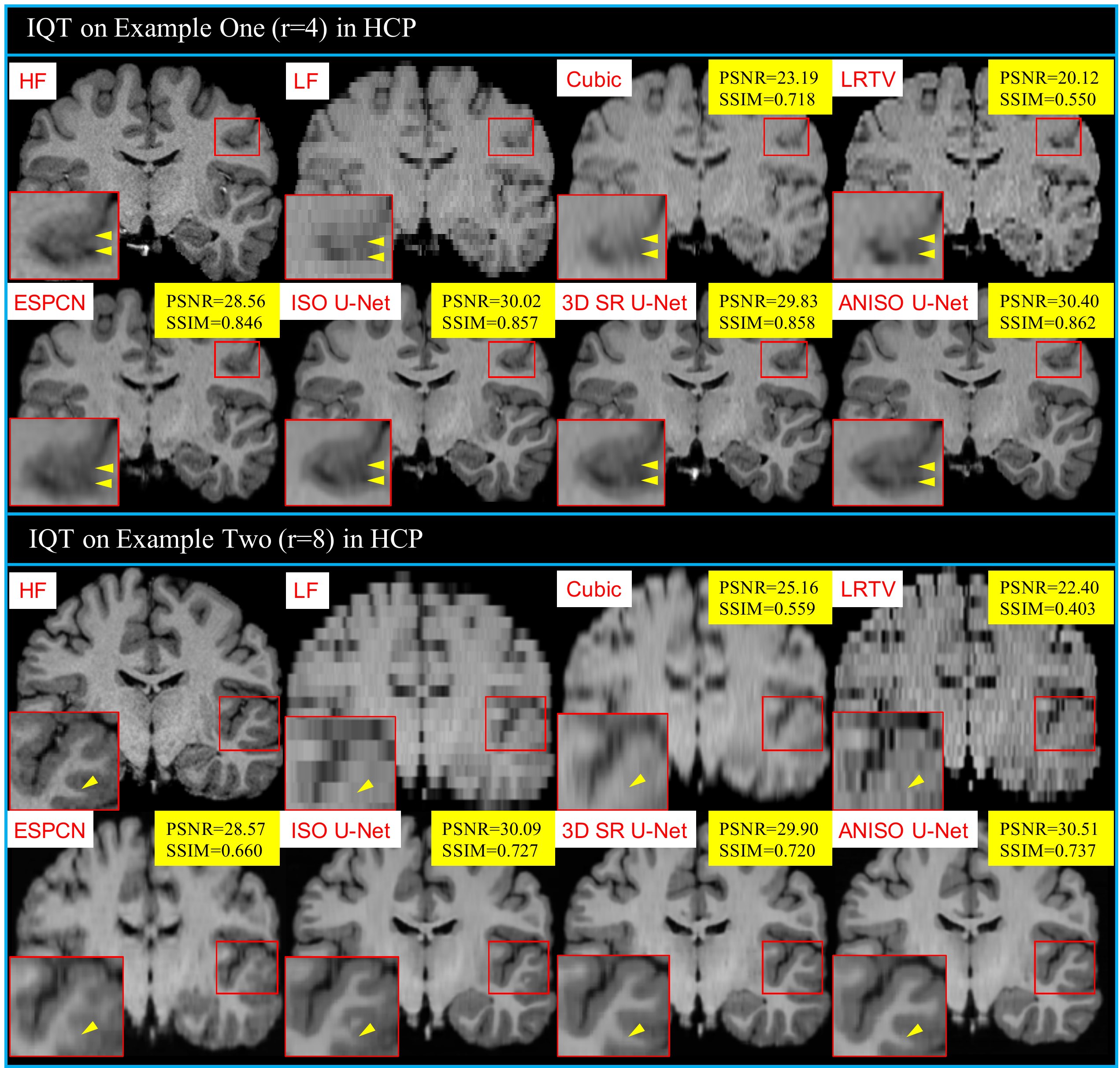}
    \caption{Coronal visualisations of the same IQT-enhanced low-field images as in Figure~\ref{fig:3.3-sagittal}. The set of methods compared is as described in the caption of Figure~\ref{fig:3.3-sagittal}.}
    \label{fig:3.3-coronal}
\end{figure}

\begin{figure}[!t]
    \centering
    \includegraphics[width=0.48\textwidth]{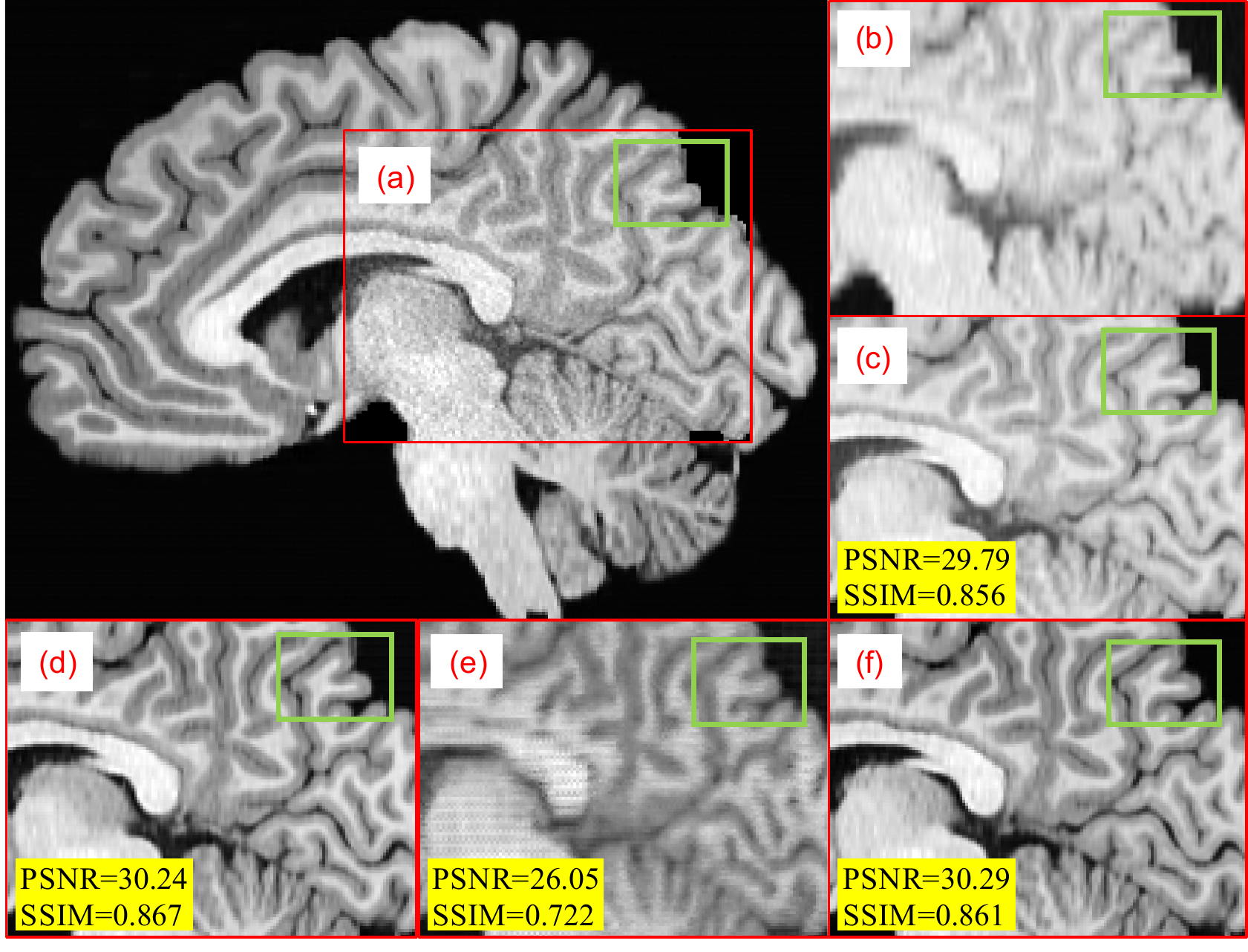}
    \caption{As Figure~\ref{fig:3.5-vis-cor} but a different example using a sagittal view. The highlighted regions corresponds to (a) the high-field image patch, (b) the low-field image patch, RC1 (c) without and (d) with NORM, RC60 (e) without and (f) with NORM are shown in the rest panels. Once again the normalisation step leads to stronger GM-WM contrast more consistently.}
    \label{fig:3.5-vis-sag}
\end{figure}

\begin{figure}[t]
    \centering
    \includegraphics[width=0.48\textwidth]{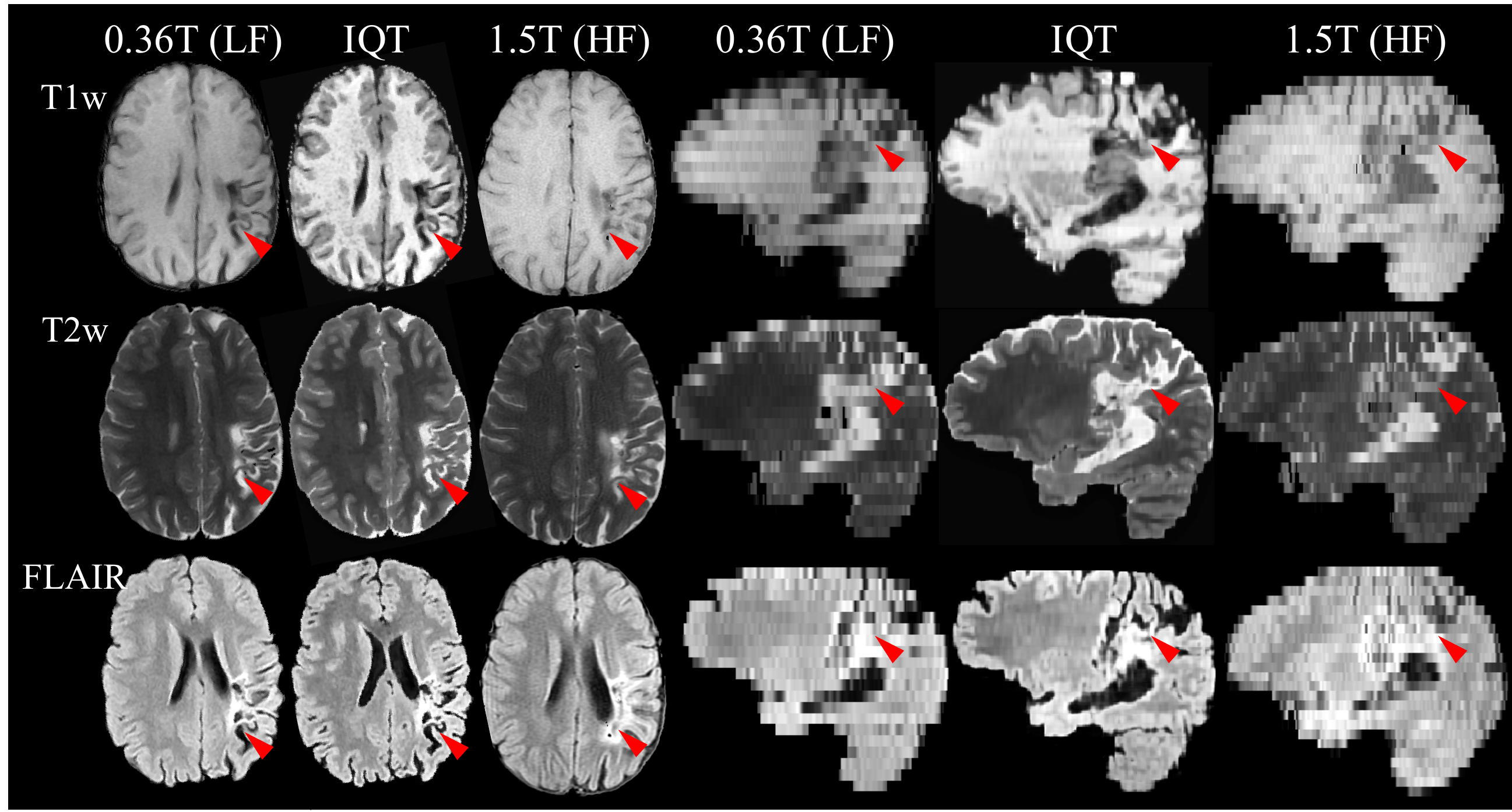}
    \caption{IQT-enhanced T1w, T2w, and FLAIR images with $r=8$, obtained from real 0.36T low-field (LF) images of a pathological subject in the \textit{LF19} data set. Red arrows point out middle cerebral artery ischemia in the perirolanding regions. All images are acquired or enhanced in 2D axial scans. The 2D axial 1.5T high-field (HF) images are provided as high-quality references. The original T1w low-field images show cross-talk and field inhomogeneity artifacts, but they were corrected during IQT pre-processing. }
    \label{fig:iqt-lesion-n19-012}
\end{figure}

\begin{figure}[!t]
    \centering
    \includegraphics[width=0.48\textwidth]{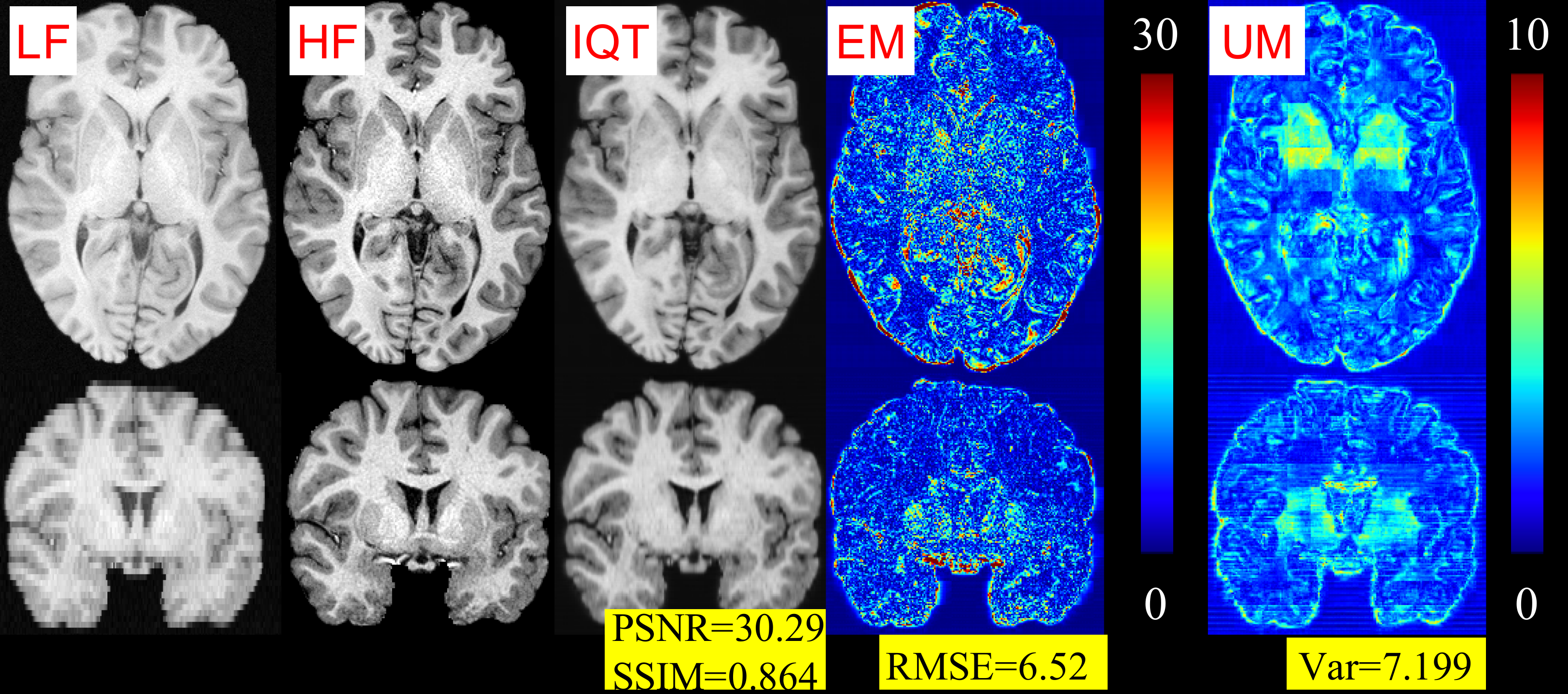}
    \caption{Low-field (LF), high-field (HF), IQT, error map (EM) and uncertainty map (UM) in both axial and coronal planes. The IQT model was trained on the RC60 data with NORM from Section~\ref{sec:results_training_contrast_norm}. The metric for EM is root mean squared error (RMSE).}
    \label{fig:uq-hcp-ax-co}
\end{figure}

\begin{figure*}[!t]
    \centering
    \includegraphics[width=.9\textwidth]{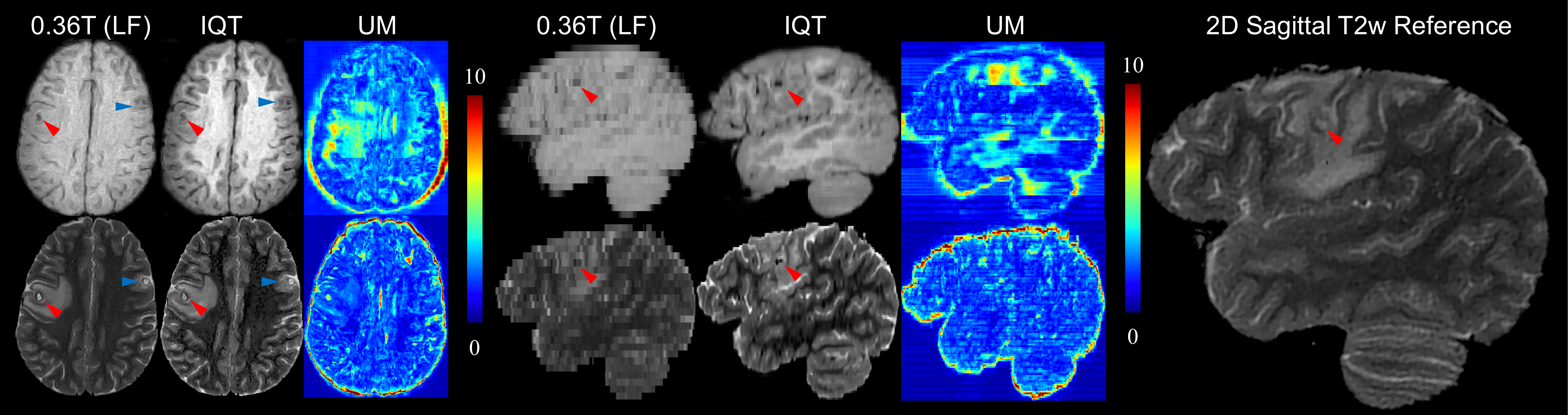}
    \caption{2D axial Low-field (LF), IQT, and uncertainty map (UM) in both axial and sagittal planes for T1w and T2w images of the same patient with cystic lesions as in Figure~\ref{fig:iqt-lesion-n17-004}. The 2D Sagittal T2w image is provided as a reference. Two lesions are pointed out in red and blue arrows.}
    \label{fig:uq-n17-ax-sa}
\end{figure*}

The following supplementary figures~\ref{fig:3.3-coronal},~\ref{fig:3.5-vis-sag},~\ref{fig:iqt-lesion-n19-012},~\ref{fig:uq-hcp-ax-co}, and~\ref{fig:uq-n17-ax-sa} are related to this article.

\section*{Data and code availability}
The IQT models are trained and evaluated on the two publicly available high-field MRI data sets, i.e. the HCP data set for T1w and T2w images~\citep{Sotiropoulos2013} and the LEMON data set for FLAIR images~\citep{babayan2019mind}. The authors do not have permission to share the clinical low- and high-field data managed by UCH Ibadan, i.e. the \textit{LF17} and the \textit{LF19} data sets. Code is available at \url{https://github.com/hongxiangharry/Stochastic-IQT}.

\bibliographystyle{model2-names.bst}\biboptions{authoryear}
\bibliography{refs}

\end{document}